%% file: main.tex
\documentclass[sigconf, review=false, anonymous=false]{acmart}

\usepackage{tabularx}
\usepackage{dcolumn} 
\newcolumntype{d}[1]{D{.}{.}{#1}}

\usepackage{subcaption}
\usepackage{draftwatermark}
\SetWatermarkText{Author Manuscript}
\SetWatermarkScale{.4}

\usepackage{todonotes}
\let\xtodo\todo
\renewcommand{\todo}[1]{\xtodo[inline,color=green!50]{#1}}

\AtBeginDocument{%
  }

\setcopyright{acmcopyright} 
\copyrightyear{}
\acmYear{}
\acmDOI{}

\acmConference[]{}{}{}
\acmPrice{}
\acmISBN{}

\begin{document}

\title{Sense of Agency in Closed-loop Muscle Stimulation}


\settopmatter{authorsperrow=3}

\author{Lukas Gehrke}
\orcid{0000-0003-3661-1973}
\affiliation{%
  \institution{TU Berlin}
  \city{Berlin}
  \postcode{10623}
  \country{Germany}
}
\email{lukas.gehrke@tu-berlin.de}

\author{Leonie Terfurth}
\orcid{0000-0001-6143-4222}
\affiliation{%
  \institution{TU Berlin}
  \city{Berlin}
  \postcode{10623}
  \country{Germany}
}
\email{leonie.terfurth@tu-berlin.de}

\author{Klaus Gramann}
\orcid{0000-0003-2673-1832}
\affiliation{%
  \institution{TU Berlin}
  \city{Berlin}
  \postcode{10623}
  \country{Germany}
}
\email{klaus.gramann@tu-berlin.de}

\renewcommand{\shortauthors}{Gehrke et al.}

\begin{abstract}
\input{input/0_abstract}
\end{abstract}





\maketitle

\input{input/2_introduction}
\input{input/3_related_work}

\input{input/5_user_study_and_methods}

\input{input/6_results}
\input{input/7_discussion}











\begin{acks}
\input{input/9_acknowledgements}
\end{acks}

\bibliographystyle{ACM-Reference-Format}
\bibliography{main}


\end{document}

%% file: input/0_abstract.tex

To maintain a user's sense of agency (SoA) when working with a physical motor augmentation device, the actuation must align with the user's intentions. In experiments, this is often achieved using stimulus-response paradigms where the motor augmentation can be optimally timed. However, in the everyday world users primarily act at their own volition. We designed a closed-loop system for motor augmentation using an EEG-based brain-computer interface (BCI) to cue users' volitional finger tapping. Relying on the readiness potentials, the system autonomously cued the finger movement at the time of the intent to interact via electrical muscle stimulation (EMS). The prototype discriminated pre-movement from idle EEG segments with an average F1 score of 0.7. However, we found only weak evidence for a maintained SoA. Still, participants reported a higher level of control when working with the system instead of being passively moved.


%% file: input/2_introduction.tex
\section{Introduction}

Advances in hardware that augment a user's physical actions have reignited dreams of overcoming human limitations, recovering lost abilities and simplifying skill acquisition~\citep{Goto2020-mw, Kunze2017-co}. These technological advances include the miniaturization of the actuating hardware to wearable form factors and the direct sensing and stimulation capabilities of neural interfaces. Especially due to these characteristics, recent perspectives promote a change of the computing era from human-computer \textit{interaction} to \textit{integration}~\cite{Mueller2020-dl}. One key change in perspective is, that \textit{integrated} users share agency with the computing machinery to execute tasks. This is a critical distinction as \textit{integration} technologies are designed to directly influence people’s bodies, their actions, and the resulting action outcomes~\cite{Cornelio2022-aq}.

Besides body and outcome augmentation, action augmentation has been defined as the case where a ``system assists the user’s action to produce the intended outcome.''~\cite{Cornelio2022-aq}. Such action augmentations can be realized purely on a software integration level, for example by an AI pair programmer. When it is designed to happen on a hardware level, further challenges emerge, specifically due to shared agency, which in this case means handing over control of one's own body. Unfortunately, such augmented users often report dissociative experiences, frequently disrupting their sense of agency (SoA)~\citep{Gilbert2017-ze, Gilbert2019-uc}.

\begin{figure}[!h]
    \centering
    \includegraphics[width=\columnwidth]{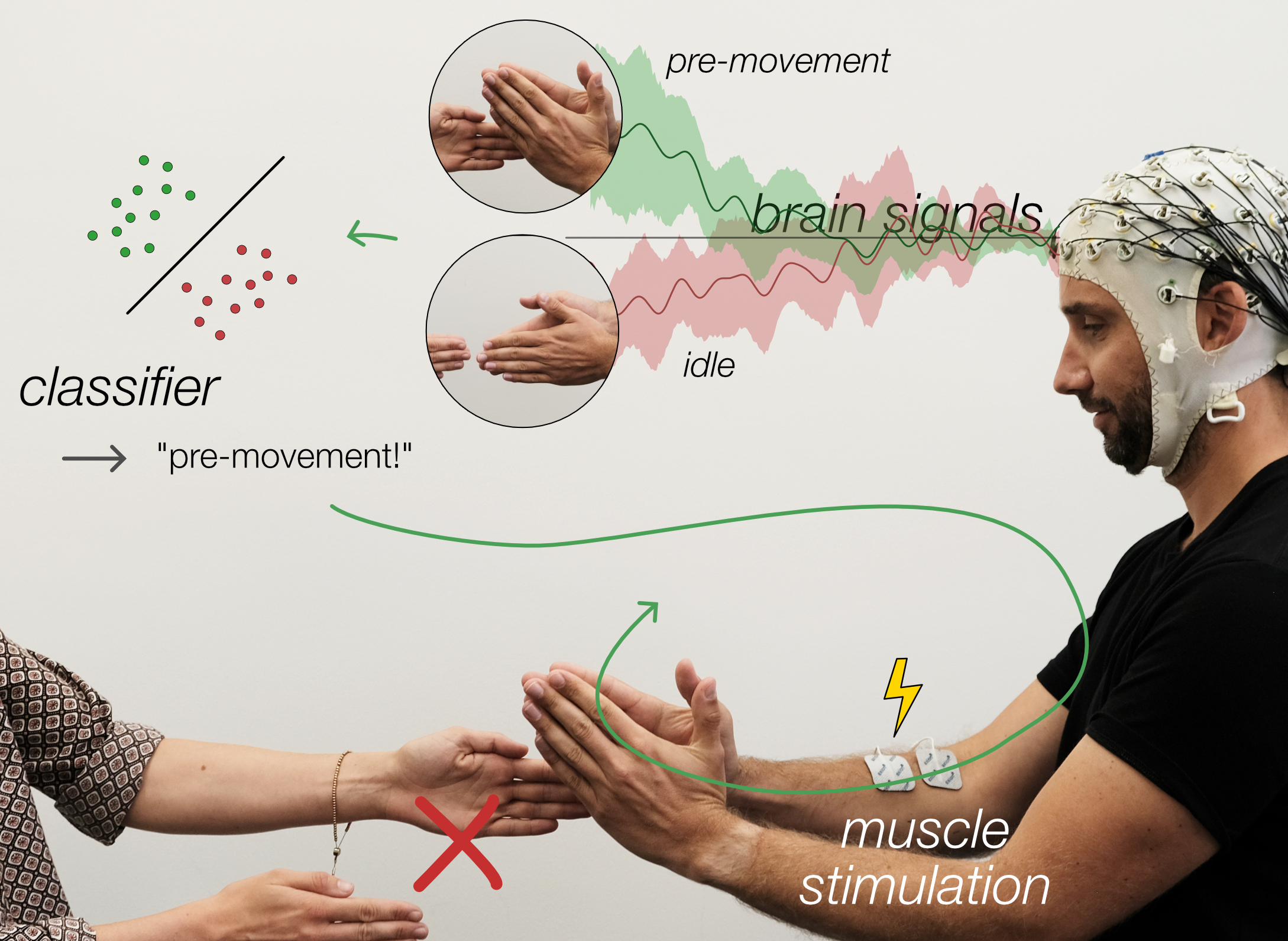}
    \caption{We propose an augmentation system that aligns with the user's agency. When participants feel the spontaneous urge to move, readiness potentials (RPs) are picked up in the user's brain signals. A brain-computer interface (BCI) then predicts the user to be in either an idle (red) or pre-movement (green) state. In the latter case, electrical muscle stimulation (EMS) is triggered and moves the user's hand. Image taken with consent from participant.}
    \label{fig:teaser}
\end{figure}

Having an SoA means experiencing control over our own voluntary actions, instead of them feeling as randomly happening to us, for a recent review see~\cite{Wen2022-bm}. It has been shown that users are more likely to feel engaged and satisfied with an interaction, and are more likely to trust a system the more they experience SoA~\citep{Berberian2012-do, Miller2007-rb}. Hence, a key challenge to drive the adoption of human action augmentation is to design for agency experience, so users feel as though they are in the ``driver's seat'' once again.

With this paper, we strive towards an `agency-aligned' augmentation system. We developed a brain-computer interface (BCI) that establishes a fast communication channel between a user's brain signals and a physical end effector, see figure~\ref{fig:teaser}. The closed-loop augmentation system (from hereon referred to simply as \textit{system}) controls the user's muscles at the time of their \textit{intent to interact}, as measured through readiness potentials (RP) manifesting in the user's electroencephalogram (EEG)~\cite{Schurger2021-vp, Schultze-Kraft2016-bx, Schultze-Kraft2021-cu}. Ultimately, such an augmentation can cue the user's movements, increase their strength, and might also preempt their action, i.e., increase their speed.


To control the interaction in real-time, an RP-based classifier distinguished between two user states: \textit{idle}, reflecting the absence of an intent to act, and \textit{pre-movement}, indicating the presence of an intent to act. During \textit{idle}, participants were passively looking at a fixation cross. Instead, during \textit{pre-movement}, participants were instructed to voluntarily initiate a tap on a touchscreen whenever they felt the urge to do so. Previous work has indicated that the RP emerges during formation of conscious intention and is specific to voluntary action~\cite{Schultze-Kraft2020-rm, Travers2020-hf, Pares-Pujolras2019-ll}.

Upon predicting a \textit{pre-movement} state, the system augments the user's action, potentially even preceding their voluntary motor command. This augmentation moves the ring finger in accordance with the user's intention to act. We achieved the movement by leveraging electrical muscle stimulation (EMS) applied to the user's forearm flexor muscle. To get a first glimpse of how our system can maintain agency in augmented users, we conducted a user study using both quantitative and qualitative research methods. In short, we explored whether aligning the physical impact on the user's body with their intention to move, preserved their SoA.

%% file: input/3_related_work.tex
\section{Related Work}
Our research draws inspiration from neuroscience and from engineering work on BCIs as well as on physical user augmentation. In order to situate our findings, we briefly review the literature on SoA specifically focusing on what it means to act at one's own volition, being a passive observer, and when acting integrated with a technology.

\subsection{Theories on Sense of Agency}
The most widely used theory on how the SoA arises is the \textit{comparator model}~\cite{Blakemore2002-dj, Frith2000-ch, Frith2006-sc}: When we act at our own volition and intentionally perform an action, the brain generates sensory predictions about the action outcome. These predictions are constantly compared to the actual sensory data available during the execution of the action. These include continuous signals such as proprioceptive and visual monitoring of the ongoing movement as well as higher level predictions about the semantic outcome of the action~\cite{Clark2013-ah, Haggard2003-ff, Haggard2017-uv}. If no sensorimotor incongruency arises, and further, the brain attributes subjective causality over the action outcome, SoA manifests. 

In the simple case of pressing a key on a piano, the finger movement is constantly compared to the predicted proprioceptive feedback. Subsequently, the tone generated by the key press is evaluated against auditory predictions. On a semantic level, these predictions may be in reference to whether the tone loudness corresponds to the velocity of the key press or whether the tone is in-key or out-of-key, and in general aligns with the subjective goal of the keypress~\cite{Pangratz2023-ew}. If these predictions -- based on the intended movement and its expected outcome -- explain the sensory data available, agency is experienced.

In human-computer interaction (HCI), these constructs are often categorized in slightly different terms. \textit{Pre-reflective} is used to describe `early', implicit, experience of agency, such as when matching proprioceptive predictions about finger movements. At higher levels of the cognitive hierarchy, \textit{reflective}, i.e. conscious, experience refers to matching semantic predictions about action outcomes~\cite{Danry2022-xk, Cornelio2022-aq}. 

In order to measure SoA, both explicit and implicit methods have been developed. Explicit methods directly query participants to report their subjective experiences using questionnaires. Items such as ``It felt like I was in control of the hand I was looking at''~\cite{Haggard2002-sz} or ``Indicate how much it felt like moving the joystick caused the object on the computer screen to move''~\cite{Ebert2010-lu}, query either the \textit{pre-reflective} action -- or the \textit{reflective} outcome evaluation~\cite{Moore2012-dk}. In most cases such questionnaires aim at a higher-level, reflective, judgment of agency.

On the other hand, implicit methods are often used to query low-level pre-reflective sensory predictions that are not consciously perceived~\cite{Moore2016-ub, Limerick2014-un, Moore2012-ic}. Seminal work in neuroscience has described one effect of SoA as a bias in the perception of action \textit{outcome}: Intentional binding paradigms state that when a button press is followed by a -- delayed -- outcome, e.g. a sound, participants mentally compress the delay~\cite{Haggard2002-sz}. In the theories original formulation, this temporal compression was assumed to only occur following movements that were intended: The action outcome is mentally \textit{bound} to the intention. To reduce uncertainty about the binding, the brain `explains away' the excess delta, compressing the action-outcome delay.

More recently it has been shown that this `temporal binding' also manifests when participants are merely a bystander witness in action-outcome scenarios~\cite{Suzuki2019-pi, Gutzeit2023-ei}. Among others, Suzuki et al.~\citep{Suzuki2019-pi} showed that temporal binding manifests as an effect of multisensory causal binding unrelated to intention or agency, e.g. binding effects were shown in scenarios where one is witnessing a replay of one's own earlier actions. Hence, it remains interesting to see how such binding manifests in \textit{integrated} user's.

As opposed to acting at one's own volition using one's own body, movement augmentation hardware allows moving a user's body even without their intention. Today, there are three main technologies to physically augment users' actions: Through the use of mechanical actuators, i.e., exoskeletons, a user's body can be moved by applying forces to the extremities~\cite{Kuhn2013-ls}. Another possibility is to stimulate the brain directly~\cite{Haggard2002-sz}, so the stimulation causes a motor response, for example by using transcranial magnetic stimulation (TMS). Lastly, electrical muscle stimulation (EMS) makes the user's extremities move by sending current into their muscle-activating nerves. Irrespective of the method applied, these technologies allow to move a user's body without the user having generated any predictions about the movement and its outcome. However, proprioceptive, visual, and other signals indicate that one's own body is moving. Hence, re-afference signals are present without an efference command and copy. Thus, concerning the comparator model, a prediction error will arise, negatively impacting SoA~\cite{Wen2020-dk}.

\subsection{Controlling Actuated Haptic Experiences}

Experimental setups to investigate new `on-body' augmentation technologies that aim to preserve the user's SoA, frequently use highly controlled `stimulus-response' paradigms. For example, scenarios where participants are instructed to tap on a touchscreen in response to a presented stimulus on the screen. Here, participants' behavior can be predicted with very high certainty to follow the presented stimulus, and estimating their reaction time is very accurate. In such controlled scenarios, the timing of an action augmentation device can be tuned to be near optimal. Hence, \textit{pre-empting} the user's motion can be designed to fall in line with their intention to move, thereby maintaining SoA. Previously, ~\citet{Kasahara2019-sk} used a reaction time task in which participants had to tap a target on screen as soon as it appeared and subsequently rate their SoA. They showed that in such a scenario, user's actions can be pre-empted and that a pre-emption of about 80 ms best preserves agency~\cite{Kasahara2019-sk, Kasahara2021-gy}. 

This is in line with evidence from cognitive neuroscience which indicates that from around 200 ms before a voluntary movement, users are unable to ``veto'' their self-initiated movement~\cite{Schultze-Kraft2016-bx}. After this ``point of no return'' user's struggle to assign a source other than themselves to the action initiation. Here, the \textit{key} aspect for SoA in action augmentation becomes apparent: External influences on the user's body need to be in line with the user's intention to act. Crucially then, a key challenge remaining is to design systems that maintain agency when user's actions are unpredictable and where the experimenter does \textit{not} have executive control over the environment. In other words, how can a closed-loop system to deliver a \textit{natural} agency experience for users' augmented actions be designed?

\subsubsection{Using Brain Signals Reflecting the Intent to (Inter-)act for Action Augmentation}
One possible design solution is to leverage physiological signals for action augmentation. Of the possible physiological signals that can be leveraged, the EEG is very well suited because of its high temporal resolution and the non-invasive recording close to the motor command generating structures in the human brain. 

The RP, or \textit{lateralized readiness potential}, is an amplitude fluctuation in the ongoing EEG activity that has frequently been observed preceding voluntary action~\cite{Deecke1969-bl, Libet1983-qu}. The RP is reliably observed at electrodes placed over the sensorimotor cortex contralateral to the acting hand. In the extended 10-20 system for EEG electrode placement~\cite{Jasper1983-uw}, these are electrodes C3 located over the sensorimotor cortex of the left hemisphere, and C4 vice versa. However, activity observed at electrode Cz is reported most frequently as it reflects neural activity originating from the sensorimotor cortex without lateralization bias. Since the RPs' measurable onset precedes the time of participants' self-reported conscious movement intention, it has drawn much interest with respect to the debate on free will, see~\cite{Schurger2021-vp} for a recent neuroscientific perspective. However, evidence abounds for its role in action preparation. An RP is typically comprised of two stages: an early slow stage that begins up to two seconds before the actual movement and a late steep stage that starts about 400 milliseconds before movement. The first stage manifests in the pre-supplementary motor area and transfers to the premotor cortex shortly after. The second stage manifests contra-laterally in the primary motor cortex~\cite{Shibasaki2006-mt}. 

A recent study has shown that the RP is ingrained in the subconscious mechanisms preceding movements that people cannot explicitly suppress~\cite{Schultze-Kraft2021-cu}. In their study,~\citet{Schultze-Kraft2021-cu} asked participants to find a way to perform voluntary movements while keeping accompanying RP amplitudes as small as possible. After each trial they informed participants about the strength of the RP in the current trial, so participants had a feedback metric to optimize for. They found participants unable to suppress their RP. This inability to suppress the RP renders it a reliable feature for classification. For example, the RP can be detected in real-time using a brain-computer interface (BCI).~\citet{Schultze-Kraft2016-bx} demonstrated a prototype that detects RPs in participants ongoing EEG data and adapts an interface accordingly. In their study, participants were instructed to veto their self-initiated movement whenever a red dot occurred on the screen. The red dot's appearance was controlled by the BCI. Whenever an RP was detected, the red dot appeared. The authors found that participants were able to veto their self-initiated movement if the red dot appeared no later than 200ms preceding their movement onset. After that, participants were unable to ``overwrite'' their motor command and acted regardless of the red dot's appearance on screen.

Taken together, these findings demonstrated that the RP is a reliable signal preceding voluntary movement initiation and hence, is a well-suited candidate to base (neuro)--adaptive systems on.

%% file: input/5_user_study_and_methods.tex
\section{User study \& Methods}
In this paper, we present a prototype that uses a user's own brain signals as the control signal to a physical end effector. With a user study, we wanted to find out whether the experience of agency can be preserved during physical action augmentation when using our prototype. 

We compared three conditions of agency experience. With the first two conditions, INTENTION and INVOLUNTARY we queried users at the two edges of agency experience. In INVOLUNTARY, participants had no intention to move and no control over their movement, much like in a bystander scenario. On the other hand, in INTENTION, participants were acting as they would in their day-to-day lives, fully in control and with volition. We then investigated in a third condition where our AUGMENTED prototype sits on the spectrum between INTENTION and INVOLUNTARY, in other words, whether it preserved agency and to what degree. To answer our question, we employed a mix of qualitative and quantitative methods, including a psychometric test of intentional binding, one standardized question, a qualitative (phenomenological) exit interview, and an exploratory EEG analyses of prediction errors following muscle stimulation.


\subsection{Participants}
Eleven participants (M = 29.9 years, SD = 4) were recruited from our local institution and through the institute's online participant pool. Participants were compensated with course credit or 12 Euro per hour of study participation. Prior to their participation, they were informed of the nature of the experiment, recording, and anonymization procedures and signed a consent form. The experiment was approved by the local ethics committee of the Department of Psychology and Ergonomics at the TU Berlin (Ethics protocol approval code: BPN\_GEH\_2\_230130). One participant had to be excluded from further data analyses due to significant deviations from the instructions in the execution of the task. Precisely, they did not initiate their movements at their own volition but rather immediately at the onset of each trial, thereby violating the task instruction of waiting 2-3 s after trial onset, see task description below for more detail.

\subsection{Apparatus}
\begin{figure}[!h]
    \centering
    \includegraphics[width=\columnwidth]{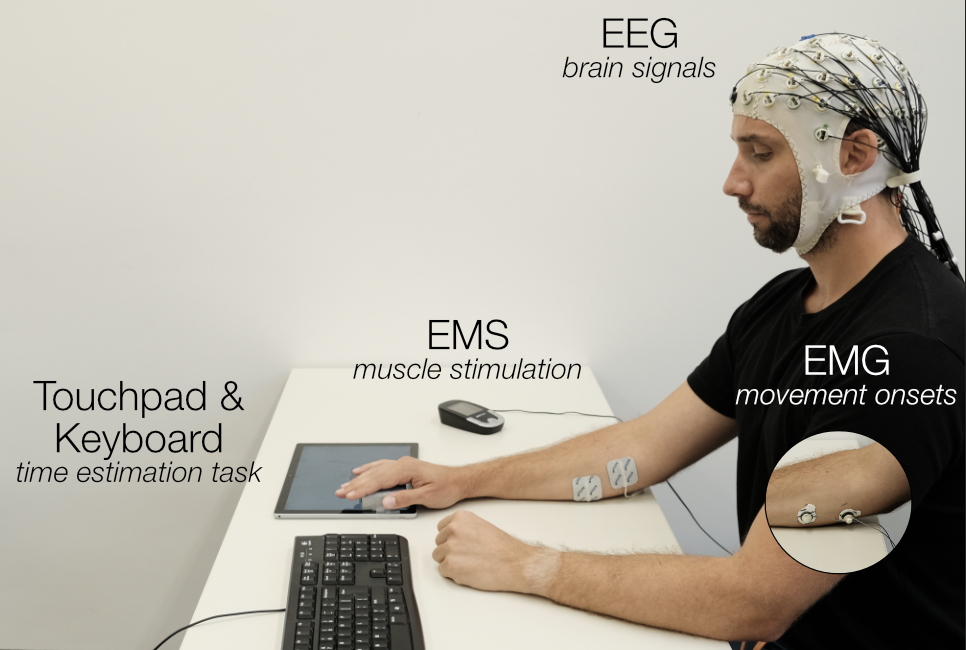}
    \caption{Experimental setup of measurement-- and input devices (image with consent from participant).}
    \label{fig:setup}
\end{figure}

The experimental setup, depicted in Figure~\ref{fig:setup}, comprised: (1) a 1-channel Electromyography (EMG) device, (2) a 64-channel EEG system, (3) a medically-compliant EMS device connected to two electrodes worn on the forearm, and (4) a tablet to run the experiment and collect behavioral responses.

\indent\textbf{(1) EMG Recording.} EMG data was recorded from 1 bipolar channel using the BrainAmp ExG amplifier (BrainProducts GmbH, Gilching, Germany). The two electrodes were placed above the flexor digitorum profundus with a reference electrode located on the wrist bone. EMG data was collected in synchrony with the EEG data through BrainProducts' BrainVision Recorder. EMG data was only recorded for the first experimental condition to obtain labels for classifier training. After that, the EMG electrodes were changed for EMS electrodes.

\indent\textbf{(2) EEG Recording.} EEG data was recorded from 64 actively amplified Ag/AgCl electrodes referenced to electrode FCz in an actiCap Snap cap using BrainAmp DC amplifiers from BrainProducts. Electrodes were placed according to the extended international 10–20 system \cite{Jasper1983-uw}. One electrode was placed under the right eye to provide additional information about eye movements (vEOG). After fitting the cap, all electrodes were filled with conductive gel to ensure proper conductivity, and electrode impedance was brought below 10k$\Omega$ where possible. EEG (and EMG) data were recorded with a sampling rate of 250 Hz. 

We used LSL\footnote{https://github.com/sccn/labstreaminglayer} to make the data streams available in the network and synchronize the recordings of EEG/EMG data and an experiment marker stream that marked sections of the study procedure.

\indent\textbf{(3) Electrical Muscle Stimulation.} We actuated the ring finger via EMS, which was delivered with two electrodes attached to the participants' flexor digitorum profundus muscle. We utilized the flexor digitorum profundus since we found that we can robustly actuate it without inducing unintended motion of neighboring fingers. This finger actuation was achieved via a medically compliant battery-powered muscle stimulator (TENS/EMS Super Duo Plus, prorelax, Düren, Germany). The EMS system's output was controlled by flipping a solid state relay (silent) connected via an Arduino Uno (Arduino, Monza, Italy) to the experiment computer. The EMS was pre-calibrated by the participant to ensure a pain-free but effective stimulation and robust actuation leading to an `immediate' tap on the touchscreen after actuation. To ensure a comfortable experimental experience so that participants relaxed their arm musculature as much as possible, a custom built hand rest (support device) was placed on top of the touchscreen for participants, see figure~\ref{fig:setup}.

\indent\textbf{(4) Experiment Presentation and Collection of Behavioral Responses.} An Acer Group (Acer Inc, Taipeh, Taiwan) tablet was used to present the task to participants and record their behavioral responses. In addition to the tablet, we used an INVOLUNTARY keyboard to allow users to input their timing judgments.

\subsection{Experimental Task, Design and Procedure}
Participants performed a simple tapping task in three conditions with 75 trials each: INTENTION, INVOLUNTARY, and AUGMENTED

\begin{figure}[h]
    \centering
    \includegraphics[width=\columnwidth]{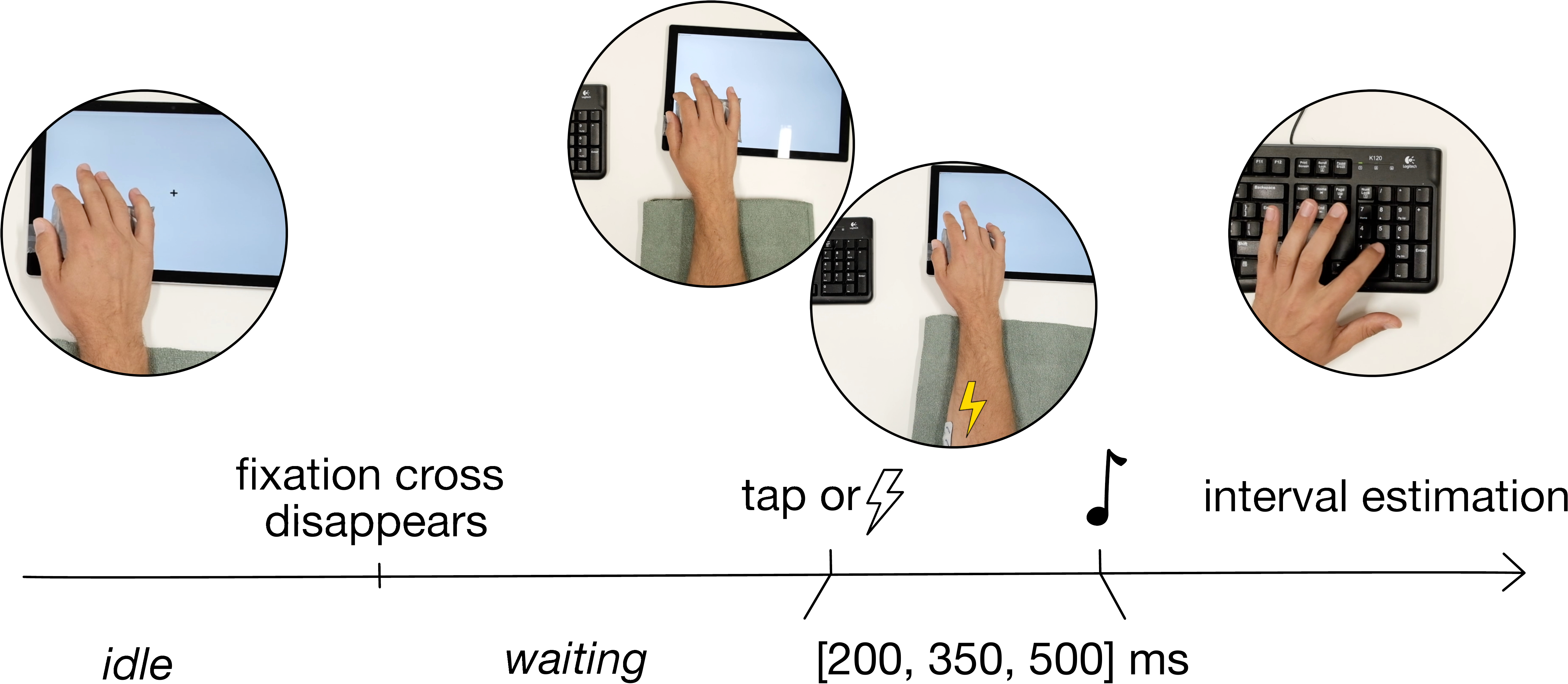}
    \caption{Interaction flow depicting one trial in our touchscreen tapping task.}
    \label{fig:progression}
\end{figure}

\begin{figure*}[t]
    \centering
    \includegraphics[width=\textwidth]{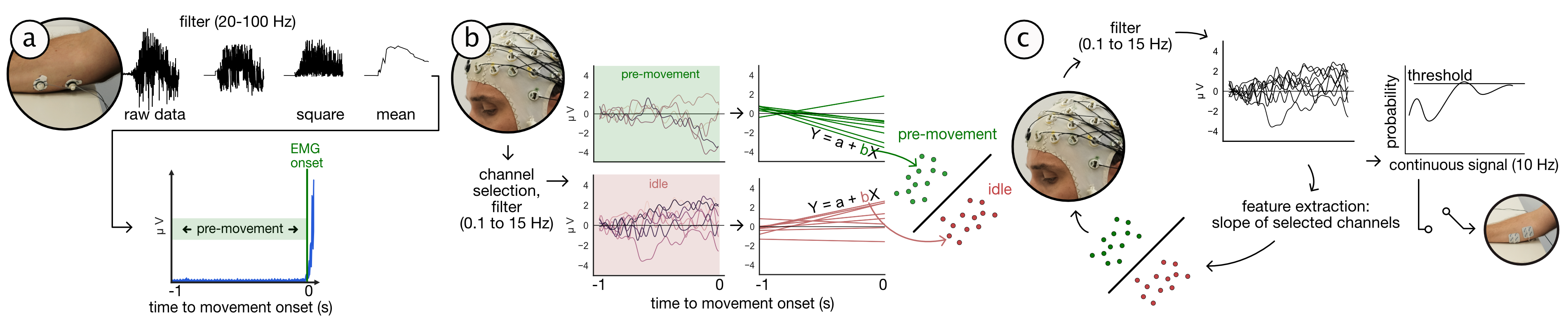}
    \caption{Diagram summarizing(a) signal processing and event labeling from muscle activity, (b) extraction of linear slope features from labeled brain signals, and (c) real-time application of the BCI gating the muscle stimulation hardware.}
    \label{fig:method}
\end{figure*}

\indent\textbf{INTENTION.} The task was as follows: (1) a fixation cross appeared on a tablet screen and participants were instructed to rest and wait until it disappeared; (2) they were instructed to wait for a brief moment (2 to 3s), before (3) initiating their movement and tap the screen, see figure~\ref{fig:progression}. In line with the literature on the origin of the RP generating process, they were told ``to avoid pre-planning the movement, avoid any obvious rhythm across trials, and to press when they felt the spontaneous urge to move''~\cite{Schultze-Kraft2021-cu}. (4) After the screen was tapped, a tone was played at a pseudo-random delay of 200, 350, or 500ms. Participants were now asked to estimate the delay, typing in their answers on a number pad of an attached keyboard. After confirming their answer by hitting the return key, the next trial started.

During INTENTION, participants were equipped with EMG sensors instead of EMS electrodes. Participants' reaction times in INTENTION were used to select stereotypical reaction times for the INVOLUNTARY condition. At the end of INTENTION, EMG electrodes were exchanged for EMS electrodes at the identical location on the forearm.

\indent\textbf{INVOLUNTARY.} The task structure was identical to INTENTION, however, participants were now instructed to hold and wait for the muscle stimulation to move their finger thereby eliciting the screen tap. The timing for the EMS trigger was taken by randomly choosing a time between the 5th and 95th percentile of their actual individual reaction time in the INTENTION condition. 

\indent\textbf{AUGMENTED.} The task and instruction were identical to INTENTION with one additional instruction: ``you will now work \textit{with} the system''. During AUGMENTED, the muscle stimulation hardware was controlled by the BCI. The classifier was set to active after the fixation cross disappeared and until a screen tap was registered. Hence, the muscle was not stimulated at other times during a trial, so as not to interfere with participants typing in their time estimation response.

The order of the conditions was not pseudo-randomized since training data obtained in INTENTION was required for both INVOLUNTARY and AUGMENTED. Furthermore, AUGMENTED was always the last condition, thereby allowing for a prolonged interview immediately after the experience with the prototype.

\subsection{Brain-Computer Interface}\label{BCI}
The data obtained in INTENTION was used to train the BCI. For processing EEG and EMG data from INTENTION, we utilized the EEGLAB~\cite{Delorme2004-sn} toolbox with wrapper functions from BeMoBIL-pipeline~\cite{Klug2022-lc} running in the MATLAB (The MathWorks Inc. Natick, MA, USA) environment. First, to generate behavioral labels at a high temporal resolution for the EEG-based classifier, we leveraged EMG data from the flexor digitorum profundus. EMG amplitudes were band-pass filtered from 20 to 100 Hz and subsequently squared. Next, to label the time of movement onset, the EMG data was averaged across trials for the second preceding the screen tap. From this averaged data, the first sample where the EMG amplitude exceeded the 95th percentile was selected as the time of movement onset, see~\ref{fig:method}a.


Two event classes were then defined as follows: \textit{pre-movement} from -1000 to 0 ms preceding the (EMG detected) movement onsets and \textit{idle}, a one-second data segment between trials where participants were looking at a fixation cross.

\subsubsection{Preprocessing EEG and Selecting discriminative channels}\label{eeg_methods}
The EEG data was band pass filtered from 0.1 to 15 Hz. In the first step to prepare the EEG data for classifier training, noisy data segments were rejected. To this end, the EEGLAB function `autorej' was used, keeping default parameters. A trial was excluded if \textit{either} data from the \textit{idle} or \textit{pre-movement} class was rejected.

Then, to decrease the dimensionality of the EEG data, we selected discriminative channels following an approach outlined by~\citep{Schultze-Kraft2021-cu}. First, for both \textit{pre-movement} and \textit{idle} the mean signal in the last 100 ms of each segment was subtracted from the mean signal in the first 100 ms of the segment. The resulting value thus indicated how much the signal had changed in the course of the 1 s segments. In line with the literature, there should be a change over time in the \textit{pre-movement} segment but not in the \textit{idle} segment. Then, all channels were sorted (1) in descending order by the signal difference in the \textit{pre-movement} segments and (2) in ascending order by the signal difference in the \textit{idle} segments. With respect to the ability of the system to discriminate between movement intention and the absence of movement intention, an ideal channel should show a strong change in signal over the course of a \textit{pre-movement} segment but no difference during an \textit{idle} segment. The ranks of the two criteria were joined by summation and the resulting order was saved for later use in training and real-time application of the classifier. Lastly, channels C3, C4, and Cz were moved to the top of the order for every participant as these are most frequently reported in studies on the RP.

\subsubsection{Training of EEG classifier}
To classify EEG data, a linear discriminant analysis (LDA) with shrinkage regularization (automatic shrinkage using the Ledoit-Wolf lemma~\cite{Ledoit2004-bi}) was trained for each participant individually. As single-trial features, the (linear) slope coefficient was obtained for both \textit{idle} and \textit{pre-movement} segments by fitting a linear regression. The classifier was then trained using the slope coefficient of the selection of the most discriminative channels as features, generating a feature vector in the dimension of channels that were kept for classification, see~\ref{fig:method}b. We purposefully constrained the dimensionality of the feature vector to avoid over-fitting and decrease the computational load.

Using scikit-learn~\cite{Pedregosa2012-sj}, the LDA was cross-validated (using 5-folds) for a grid search from 6 to 20 channels with a step size of adding 2 channels. Following this grid search, the cross-validation with the highest accuracy determined the number of channels that were kept for training the model for real-time application. Ultimately, to determine the threshold at which, during the real-time application, the classifier would trigger the EMS, we computed the Receiver-operator characteristic (ROC) and from this selected the threshold at 15 \% false positive rate.

\subsubsection{Real-time application and EMS control}
During real-time application, the EEG data was buffered for the last second for the selected discriminative channels. The data was band-pass filtered analogously to the training data from 0.1 to 15 Hz. Next, the slope was computed. This procedure ran at an update rate of 10 Hz, hence every 100 ms a new prediction was obtained from the classifier. To smooth the prediction output to reduce false predictions due to unlikely peaks, the predicted probability for the \textit{pre-movement} class was smoothed by averaging the current and the preceding prediction with a weighting (weight of .3 for the preceding, and .5 for the current prediction). The weights were obtained through trial-and-error during piloting. Then, with a 10 Hz update rate, this smoothed probability, as well as the predicted class for the current frame were gating the EMS switch: when the probability exceeded the threshold and the currently predicted class was \textit{pre-movement}, the switch was opened for 0.5 s, see figure~\ref{fig:method}c.


\subsubsection{Post-hoc evaluation of real-time performance}
To not break the user's focus on the task at hand, (subjective) labels were not obtained during the real-time application in AUGMENTED. In other words, participants were not directly queried, e.g. via questionnaires, to judge whether the stimulation in the current trial was in line with their intention or not. Without these labels, a post-hoc analyses of the binary classifiers performance was not possible, i.e. to ascertain for example false positive stimulations. However, we explored an alternate approach to estimate (subjective) labels that does not break the user's task immersion. We investigated prediction errors in response to the movement onset through event-related potentials (ERPs) at fronto-central electrode FCz. Previously, prediction error ERPs at, among others, electrode FCz have been shown to be one suitable candidate feature to detected breaks in (task) immersion, such as when perceiving glitches in VR~\cite{Gehrke2019-og, Gehrke2022-kz, Gehrke2024-xq, Terfurth2024-kh, Si-mohammed2020-ru}. In BCIs, these ERPs are frequently leveraged to correct system errors~\cite{Zander2016-ed}. In their work,~\citet{Zander2016-ed} demonstrated a BCI to decode a user’s intended cursor movement direction on a 6 × 6 grid. The system regularly probed the user by observing the EEG response to random cursor movements. How severely the random dot movement violated the user’s intention was directly reflected in the ERP activity. We hypothesized to find similar ERP signatures when the stimulation misaligned with the user's prediction.

Since this analyses was conducted post-hoc, it allowed for more signal processing in comparison to the feature extraction from INTENTION when participants were waiting to start the next block. With the goal to best recover the prediction error ERP, we again applied `BeMoBIL-pipeline'~\cite{Klug2022-lc} wrapper functions of EEGLAB~\cite{Delorme2004-sn}: After removing non-experiment segments at the beginning and end of the concatenated recording from all three conditions, EEG data was re-sampled to $250~Hz$. Next, bad channels were detected using the `FindNoisyChannel' function, which selects bad channels by amplitude, the signal-to-noise ratio, and correlation with other channels~\cite{Bigdely-Shamlo2015-ds}. Rejected channels were then interpolated while ignoring the EOG channel, and finally re-referenced to the average of all channels, including the original reference channel FCz. After applying a high-pass filter at 1.5 Hz, time-domain cleaning and outlier removal were performed using AMICA auto rejection~\cite{Palmer2011-zs}. Eye artifacts were removed using the ICLabel toolbox applied to the results from an AMICA~\cite{Pion-Tonachini2019-fy}. For this, ICLabel's popularity classifier was used, meaning that all components having the highest probability for the eye class were projected out of the sensor data.

For all three conditions, ERPs were extracted from band-pass filtered (0.1 Hz to 15 Hz) activity at electrode \textit{FCz}. These ERPs were obtained from -1000 ms to 500 ms around the movement onset. Trials were excluded in line with the removal for classifier training, see section~\ref{eeg_methods}. For all muscle-controlled trials without EMS, movement onset was defined as the time of the tap on the touchscreen minus the `EMG-delay' defined in section~\ref{BCI}. For all EMS-controlled trials, movement onset was defined as EMS onset. To keep only EMS-controlled trials where the stimulation resulted in an `immediate' screen tap, first all trials were rejected where there was no screen tap in the 350 ms following EMS. Next, `extreme outlier removal' was conducted on the delta between EMS and the screen tap using Tukey's method based on the inter-quartile range ~\cite{Tukey1949-sl}. In total, 32.9 (SD = 30.3) trials were rejected on average per participant.

\subsection{Measures of Agency Experience: Intentional Binding, Question \& Interview}
After each condition participants were prompted to rate their experienced agency on a 7-point Likert scale with the statement ``It felt like I was in control of the movements during the task.'', the item was copied from~\cite{Bergstrom2022-fb}. Following AUGMENTED we interviewed users about their experience working with the system. After prompting users to recall their experience and summarize what their task had been, we set the focus to the tapping movement and asked them to ignore the time estimation task in the questions that followed. We entered the open part of the interview by asking: ``What did the system do?'' followed up by ``What was the difference between the three conditions''. After some time, and depending on their answers, we reset the focus to the AUGMENTED condition and asked ``How often was the system active?'' followed by ``What do you think caused the actions of the system?''. This was then followed up by an `open' interview in which we frequently asked `how' and `why' questions to inquire about the user's experience.

We analyzed the interviews by loosely following~\citet{Mayring2015-pp}. All interviews were manually transcribed and translated to English using DeepL\footnote{https://www.deepl.com/translator} (DeepL SE, Cologne, Germany). Before screening the texts, two experts clustered the responses into 3 clusters: First, ``Functionality'' referred to what participants attributed the source of the stimulation. Next, we clustered responses according to ``Guessed percentage'' of correct interaction, i.e., participants' estimate of how well the system was aligned with their intention. For the cluster ``Correct Interaction'', we specifically queried participants to recall the moments where the stimulation felt in line with their intention to move, then we clustered their responses into sentiments with positive and negative valence. 

\subsection{Statistical Analyses}
To confirm and demonstrate the discriminative power of the EEG features, we plotted the amplitude time course of electrode Cz between \textit{pre-move} and \textit{idle} epochs. Electrode Cz was chosen for exemplary presentation, as it is frequently reported in studies on the RP, since it captures neural activity originating from the sensorimotor cortex. Next, the slope coefficients were extracted and a paired t-test was conducted. Ultimately, to assess the classifier performance, we calculated the F1 score, i.e. the harmonic mean of precision and recall, and plotted the ROC, see figure~\ref{fig:EEG_results}a to c.

\subsubsection{Hypotheses Testing}
Prior to any statistical analyses of the intentional binding measure, outlier trials were rejected. We applied `extreme outlier removal' using Tukey's method~\cite{Tukey1949-sl} on three time intervals that well describe `regular' behavior across trials: (1) Tapping the screen in a reasonable interval after the fixation cross disappeared. An excessively short or long delay indicated that participants either tapped the screen prematurely by accident or they were checking in with the experimenter, respectively. (2) Providing a `reasonable' estimation in the intentional binding task. (3) The EMS stimulation leading to an \textit{immediate} screen tap. A long delay between the EMS trigger and the subsequent screen tap indicated that the stimulation was not strong enough in this trial to lead to muscle actuation resulting in a screen tap. Taken together, applying Tukey's method to each of these time windows and fusing the rejected trials led to the exclusion of 110 trials across all participants (M = 12.2, SD = 9.1). 

In line with the literature on intentional binding, we hypothesized that the time intervals should be underestimated for the INTENTION and the AUGMENTED conditions. This should not be the case for the INVOLUNTARY condition where users had no intention to move. Hence, when binding occurs, the intervals should be underestimated, hinting at a higher SoA. 

To test this, we fitted a linear mixed effects model with \textit{condition} (INTENTION, INVOLUNTARY, AUGMENTED) as a fixed effect and `participantID' as a random effect on aggregated data. The model was specified as `time interval $\sim$ condition + (1|participantID)' and fit using the `lme4' package~\cite{Bates2015-bh}. A test statistic was obtained by calculating likelihood-ratio tests comparing the full model as specified above against the null model `time interval $\sim$ 1 + (1|participantID)'. All parameters were estimated by maximum likelihood estimation~\cite{Pinheiro2006-bk}. We computed post-hoc pair-wise tests for `condition' corrected for multiple comparisons (Tukey method) using the emmeans package~\cite{Lenth2020-xk}


Next, we hypothesized that subjective ratings of control over the tap movement are comparable between INTENTION and AUGMENTED and lower in the INVOLUNTARY condition. Again, we fitted a linear mixed effects model with \textit{condition} as a fixed effect and participant ID as a random effect. Coefficients were assessed in the same way as for the intentional binding parameter above.

In short, we tested two main hypotheses with regard to the agency experience in our three experimental conditions: (1) participants underestimate the tone delay when acting intentionally. Adding EMS in line with participants' intention to move does not affect this underestimation. (2) The subjective feeling of control is comparable between INTENTION and AUGMENTED conditions. INVOLUNTARY should decrease the feeling of control significantly. Additionally, we report the clustered interviews anecdotally.

To analyze prediction error ERPs following the movement onset, potentially reflecting a disruptive experience impacting SoA, a linear mixed effects model was fit at each time point of the ERP using the `lme4' package~\cite{Bates2015-bh}. The model `ERP\_sample $\sim$ condition + (1|participantID)' was fit with condition reflecting trials belonging to INTENTION, INVOLUNTARY, or AUGMENTED. Prior to model fitting, all trials were aggregated per participant and condition. A test statistic was obtained by calculating likelihood-ratio tests comparing the full model as specified above against the null model `ERP\_sample $\sim$ 1 + (1|participantID)'. All parameters were estimated by maximum likelihood estimation~\cite{Pinheiro2006-bk}. P-values were corrected for multiple comparisons using \textit{false discovery rate}~\cite{Benjamini1995-cw} at $\alpha = .05$.

Since the AUGMENTED condition contains both, `EMS-controlled' trials, where the classifier detected an intent to interact, as well as `muscle-controlled' trials, where the classifier failed to detect the user's intent and the user carried out the tap without EMS, we split these two trial categories. We hypothesized that `EMS-controlled' trials in AUGMENTED \textit{differ} from trials in INVOLUNTARY. The latter should elicit a stronger prediction error than the former since no impact of volition is present in INVOLUNTARY, which could potentially moderate the ERP. Similarly, we hypothesized that `muscle-controlled' trials in AUGMENTED should \textit{not} differ from INTENTION trials. To test these hypotheses, permutation t-tests were conducted on aggregated data using MNE-python~\cite{Gramfort2013-fa} contrasting `EMS-controlled' trials in AUGMENTED with INVOLUNTARY as well as `muscle-controlled' trials in AUGMENTED with INTENTION. The tests were conducted for the time window from 150-250 ms post movement-- or EMS-onset event. In short, these tests were conducted to investigate the moderating role of the user acting at their own volition in AUGMENTED.

%% file: input/6_results.tex
\section{Results}

\begin{figure}[h]
    \centering
    \includegraphics[width=\columnwidth]{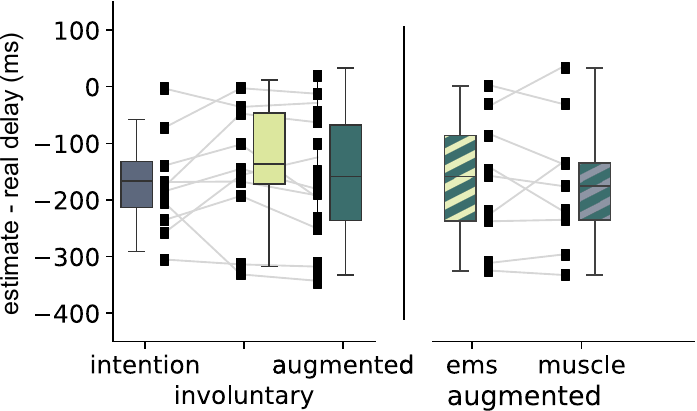}
    \caption{Difference in time estimation from real delay in intentional binding task for the three experimental conditions (left). Negative values indicate a temporal compression, i.e., temporal binding. Right side: Trials in the AUGMENTED condition split in EMS-- and self-executed trials.}
    \label{fig:ib_task}
\end{figure}

In the intentional binding task, i.e. the estimation of the time interval between tap and tone, participants generally underestimated the average real delay (350ms) in all three conditions (INTENTION M = -160.7 ms, SD = 88.1; INVOLUNTARY M = -135.3, SD = 110.3, AUGMENTED M = -162.4, SD = 113.8). The underestimation was not affected by the condition, see~\ref{fig:ib_task}.


\subsection{Subjective Ratings \& Reports}
\begin{figure}[h]
    \centering
    \includegraphics[width=.5\columnwidth]{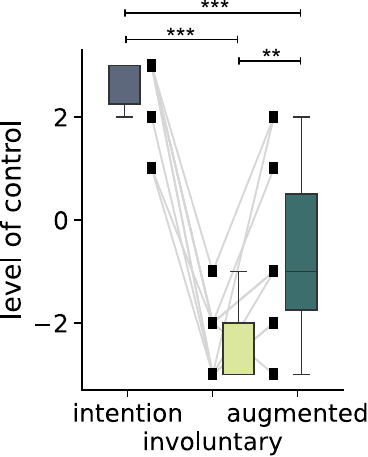}
    \caption{Subjective ratings of control across conditions. Significance labels obtained from post-hoc tests on estimated marginal means.}
    \label{fig:loc}
\end{figure}

The subjective rating of control differed between conditions (${\chi^{2}_{(2)}} = 43.7, p < .001$), see figure~\ref{fig:loc}b. Post-hoc tests revealed that participants rated their level of control higher in INTENTION compared to INVOLUNTARY ($beta = 4.8, p < .0001$), and higher in INTENTION compared to AUGMENTED ($beta = 3.1, p < .0001$). Further, higher control was observed in AUGMENTED compared to INVOLUNTARY ($beta = -2.9, p = .006$).


\paragraph{Functionality.} Eight out of the ten participants made a reference to the source of the EMS in the AUGMENTED condition. In that subgroup, most participants wondered how the system worked, one remarked, ``I can't explain how it works technically.''. Another one mentioned that ``maybe the time estimation task had something to do with it.''. They found the stimulation to be sporadic, even during their own actions, leading to a belief that it was coincidental. As one participant put it, ``I have no idea what controlled the stimulation, I think it was coincidence when stimulation occurred.''. Furthermore, participants reported that they found the timing of the stimulation to be unpredictable, with one participant noting, ``Stimulation was random in time.''. 

Some perceived the stimulation as externally triggered, yet partially responsive to their choices, resulting in a sentiment captured by, ``I think that the stimulation in the third block was partly random, but partly as if it was following my decision/choice.''. One participant noted that they felt that ``somehow the information is in my arm''. Another one briefly considered the involvement of their brain waves in controlling the stimulation, but expressed doubts about this possibility, stating, ``But I don't think that is the case.''.

\paragraph{Guessed Percentage.} 
Some participants reported that the stimulation ``sometimes overlapped with the movement, but not often,'' or, it ``came only rarely.'' Another one noted, ``In a few cases, the stimulation came when I had already started the movement.'' For several participants, this overlap between their actions and the system's response occurred infrequently, with another user noting, ``Once, in the millisecond range between my planned movement and its execution.'' However, for some participants, there were moments of near-perfect synchronization, as described by one participant, ``In 3 cases it happened that my intention to press and the impulse of the device happened simultaneously.'' or ``in 3--4 cases it happened simultaneously that i wanted to tap and the stimulation happened''. Another one noted, ``Sometimes it really happened that they overlapped. So that I just started to move, and then the device activated,'' that user estimated an overlap of ``40 \%'' while two other users said [we worked together in] ``15 \% of trials.'' These accounts collectively highlight the varied and occasionally synchronous nature of the system's timing in relation to the users' intentions.

\paragraph{Correct Interaction.} 
In terms of valence sentiments for the cases where the stimulation aligned with participants' intentions, some participants indicated that the experience was positive, for example, participants described the experience as ``rather funny'', ``weird but funny'', ``pleasant'' or ``helped me with the execution'' and ``it was more of a collective movement.'' One participant remarked ``then it was ok to experience the stimulation, but also not more than ok'' while another one noted ``It had a bit of thinking ahead to it''. Some noted that they experienced an increase in their physical strength, remarking, ``supported my strength'', ``made me type more firmly'' and ``my typing performance was increased.'' On the other hand, some participants had negative sentiments during these moments of aligned stimulation, remarking, ``felt like I was still in competition with the system'', ``it did not feel like an acting together'' and ``on a psychological level it was a loss of control.'' Another one noted, that they ``felt determined-by-others and then tried to resist the impulse. I felt excluded from the decision to tap''.

\subsection{Classifier Performance}
\begin{figure*}[t]
    \centering
    \includegraphics[width=\textwidth]{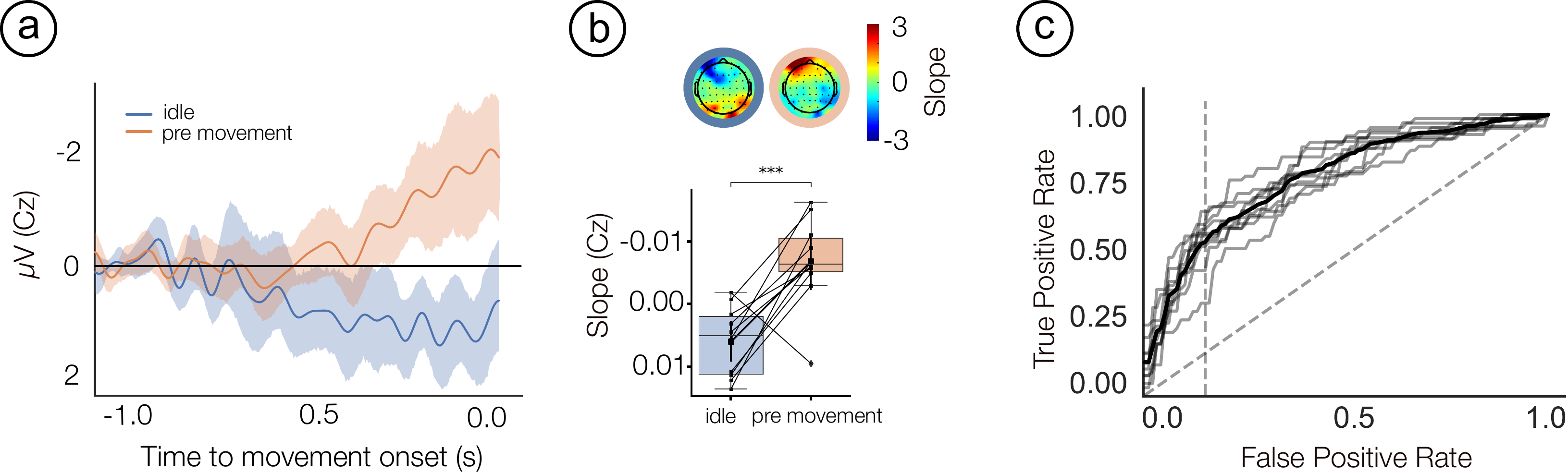}
    \caption{(a) Grand average event-related potential (ERP) at electrode Cz of pre-movement EEG data epochs preceding the last second before a movement, idle epochs of the same trial are plotted alongside but originate from a different time window, see section~\ref{eeg_methods}; (b) Bottom: Slope features for both idle and pre-movement classes at electrode Cz. Top: Scalp maps of slope values for both classes and all channels. (c): Mean and per participant ROC curves. Dotted line at 15 \% false positive rate indicate the selected threshold for the real-time application.}
    \label{fig:EEG_results}
\end{figure*}

Visual inspection of the amplitudes at electrode Cz revealed an increase in the difference between \textit{pre-movement} and \textit{idle} data segments towards the onset of the finger movement, see figure~\ref{fig:EEG_results}a. The slope feature for the exemplary channel Cz discriminates well between the two classes (${t_{(10)}} = 4.4, p < .001$), see figure ~\ref{fig:EEG_results}b bottom. The scalp maps in figure~\ref{fig:EEG_results}b top show the (color-coded) mean slope for each channel and each class. Central channels on the contralateral side to the moving finger on the right hand show a negative slope for the pre-movement class and a neutral slope for the idle class. Furthermore, differences in slope were also observed at frontal electrodes over the left hemisphere as well as parietal electrodes, were a positive slope manifested only for the idle class. 

The grid-search over channels resulted in the BCI leveraging on average 11.2 (SD = 3.6) channels. Besides channels C3, C4, and Cz, that were always included, other common channels (retained for at least 3 participants) included FT9 and AF3. The classifier cross-validation resulted in a mean F1 score of .71 (SD = .03), see figure~\ref{fig:EEG_results}c. We set the detection threshold to 15 \% false positive rate and at that rate, observed a mean threshold of 57 \% (SD = .04). Hence, on average, the classifier switched on the EMS when it predicted class \textit{pre-movement} with 57 \% probability.

\subsubsection{Post-hoc analyses of classifier performance}

\begin{figure}[h]
    \centering
    \includegraphics[width=\columnwidth]{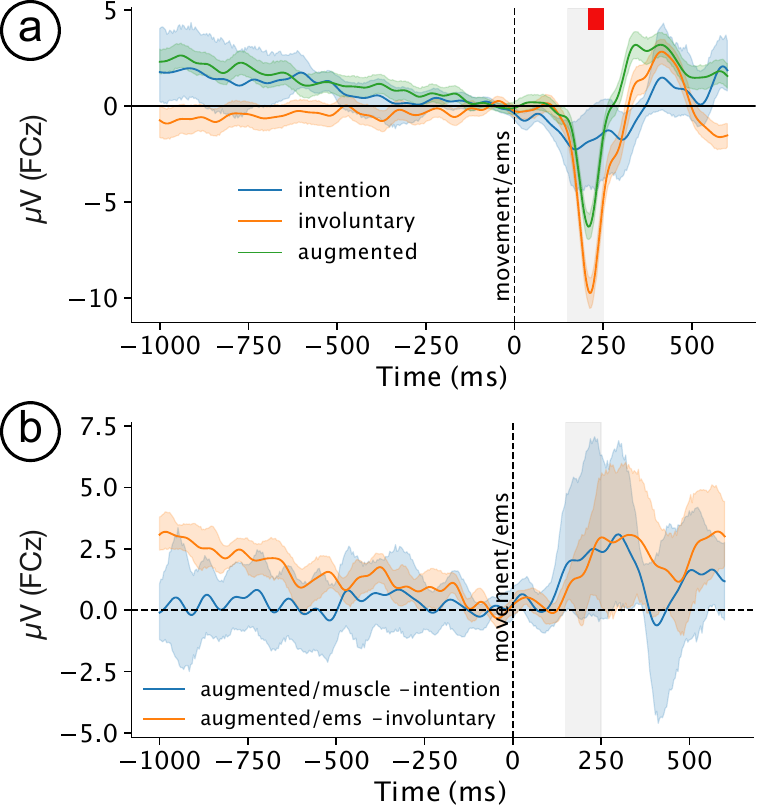}
    \caption{ERPs at electrode FCz from -1000 before, to 600 ms after the tapping movement onset. Top (a): INTENTION, INVOLUNTARY and AUGMENTED trials. Red bar on top indicate main effect at $p<.05$. Bottom (b): Difference ERPs at FCz of the `muscle-controlled' trials in AUGMENTED minus INTENTION as well as EMS trials in AUGMENTED minus INVOLUNTARY.}
    \label{fig:erp}
\end{figure}

In line with the readiness potential a negative going deflection over the last second preceding the movement onset was present in both INTENTION and AUGMENTED, but not in INVOLUNTARY, see figure~\ref{fig:erp}a. Furthermore, ERPs differed between the three conditions in the time window from 210 ms -- 250 ms following movement onset ($\chi^{2}_{(2} = 10.6, p = .03$ at 220 ms). INVOLUNTARY exhibited a strong negativity peak (strongest among the three conditions), peaking at around -10 $\mu$V at 210 ms after movement onset. A weaker peak was observed for AUGMENTED (-6 $\mu$V at 210 ms) and even more so for INTENTION (-2 $\mu$V at 150 ms).

When correcting for the influence of the EMS by subtracting INVOLUNTARY from EMS trials in AUGMENTED an INTENTION from `muscle-controlled' trials in AUGMENTED we observed no differences. Visually, it appears that trials in both INVOLUNTARY and INTENTION condition exhibit a stronger negativity in the post movement time window of interest than respective EMS and `muscle-controlled' trials in AUGMENTED, see the positive deflections peaking at 300 ms post movement onset in figure~\ref{fig:erp}b.




%% file: input/7_discussion.tex
\section{Discussion}
In this paper, we investigated if SoA can be maintained in action augmentation when the augmentation aligns with the users intention. We evaluated a BCI that controls EMS at moments of users' intention to interact. By leveraging an average of 11 EEG channels and using a simple, fast-to-compute feature, the BCI achieved a mean F1 score of about .7. 


In the user study with 10 participants we found no evidence for a disruption in intentional binding, hinting at a maintained SoA. In line with the literature, participants underestimated the delay between tap and tone in both conditions where they were instructed to act on their own volition~\cite{Moore2012-dk}. However, we also found that participants similarly underestimated the delay in INVOLUNTARY. This conflicting finding is in line with recent literature that has questioned the validity of the intentional binding phenomenon as a correlate of agency, stating that the effect may ``merely represent a strong case of multisensory causal binding.''~\cite{Suzuki2019-pi, Gutzeit2023-ei, Hoerl2020-my}. This might be especially true for cases where one's own body is completing the action while being externally controlled. Interestingly, in our scenario, proprioceptive signals and other indicators of embodied actions remain in line with the user acting at their own volition. Hence, the underestimations we found might purely follow from multisensory causal binding. The temporal compression effect might therefore not be based on inferring subjective causality following from intention but solely from causally binding sensory information. 

We do note that the underestimation appeared to be trending smaller in INVOLUNTARY as in both, INTENTION and AUGMENTED. However, the effect might be smaller than what could be proven given our sample size. Within the AUGMENTED condition trials, we observed no difference between `EMS-controlled' --and `muscle-controlled' trials, indicating an alignment with the augmentation technology in the AUGMENTED's `EMS-controlled' trials as the stimulation did not appear to disrupt participants' performance in the temporal delay estimation task.

On the level of subjective experience, we found that participants rated their level of control lower when using the system (AUGMENTED) as compared to voluntary interaction without EMS (INTENTION). This is a common finding when EMS is used for muscle control. For example,~\citet{Lopes2015-dk} reported people frequently not feeling in control in an interaction facilitated via EMS. INTENTION than might serve a very high mark for ratings following AUGMENTED and a better fitting baseline comparison would make sense. One way to better assess agency attributions when working with the system could be a comparison to earlier works that also leveraged EMS instead of our INTENTION condition where no EMS was in use. Triggering EMS based on simple heuristics (or sham) and contrasting this with BCI-controlled EMS stimulation could shed light on agency attributions exclusively explained via the BCI control. As a next step forward, stimulus-response paradigms like those of~\citet{Kasahara2019-sk, Kasahara2021-gy} could then serve as the next step-up in evaluation, with the augmentation performance, such as \textit{pre-emptive} gain, coming into focus.

From participants' qualitative reports we learned that when the stimulation aligned with participants' intentions, positive sentiments outweighed negative ones. When it did not, participants reported a negatively perceived loss of control. To situate participants comments, a differentiated look at the classification system's performance is warranted.

First, we note that our system was built using off-the-shelf, affordable, equipment to physically augment users' actions. Taken together, all technical devices to control the users' movements, i.e., the EMS stimulation device, Arduino, and switchboard, cost less than 100 Euros. While we used a 64-channel research-grade EEG system, the channel selection procedure resulted in the system ultimately using a low-density channel coverage for classification. Today, many low-density EEG devices are available on consumer markets at affordable price points, see~\citet{Niso2023-ce} for a recent summary of available wireless systems. Still, the system achieved what is considered to be a `good' F1 score of .7, and hence sometimes detected users' intention to interact.

While an F1 score of .7 may be considered `good' for many classification scenarios, when aiming to elicit a feeling of control this level of performance may likely not be good enough, see~\citet{Papenmeier2022-oi} for a review. Balancing false positive rate (FPR) and true positive rate (TPR) appropriately may prove crucial to elicit an experience of agency.

\subsection{ERPs for Data Labeling}
False negative classifications meant that the system did not trigger an EMS pulse in line with participants' intent to tap on the screen. On the other hand, false positive classification meant that an EMS pulse was sent in the absence of a true intention by the participant. Both cases, individually and jointly, had the potential to impact the user experience and erode trust in the system. 

We investigated EEG activity at electrode FCz to ascertain presence or absence of a prediction error in response to the stimulation or movement onset, see figure~\ref{fig:erp}. Ultimately, classifying the EEG here could be leveraged to approximate labels for classifier validation. The idea is that in AUGMENTED a false positive would mean that the EMS is falsely triggered and hence the movement onset would catch the participant by surprise. On the other hand a false negative means that a user starts their movement on their own without EMS, potentially equally `surprising' them, or in other words not aligning with their prediction. In line with this theory, we found a negativity affected by the trial condition in the  150--250 ms time window after movement onset, see figure~\ref{fig:erp}a. A strong negativity was present in INVOLUNTARY, where the user experienced the stimulation at unpredictable times. A less pronounced negativity was present in AUGMENTED, with only a marginal peak present in INTENTION.

To carve out the differences between the two trial categories `EMS-controlled', and `muscle-controlled' in AUGMENTED from the two \textit{anchoring} conditions, we subtracted each category from INVOLUNTARY and INTENTION accordingly, see figure~\ref{fig:erp}b for an explorative view on the ERP data.

First, we note that all `muscle-controlled' trials from AUGMENTED were false negatives, as the classifier failed to pick up on the readiness potential preceding the movement. By subtracting AUGMENTED, we observed that the pre-movement activity at electrode FCz did not differ from INTENTION. This shows that both these trial groupings exhibited a similar readiness potential at electrode FCz and the classifier failed to pick up on on it. Following the movement start, a trend emerged after 150--250 ms, in which `muscle-controlled' trials from AUGMENTED trends towards a stronger negativity than INTENTION trials, hence a positive difference, see figure~\ref{fig:erp}b blue line. This may indicate that participants always expected an EMS pulse in AUGMENTED and carrying out the movement without EMS \textit{support} violated their prediction. 

In the `EMS-controlled' trials in AUGMENTED, both false positive and true positive classifications overlap. In the contrast with INVOLUNTARY, the readiness potential which lead towards a positive classification outcome was visible in the second preceding the movement, see the negative going deflection in figure~\ref{fig:erp}b orange line. In the 150--250 ms time window after movement onset, a similar trend as described above was visible. The slight positive bump was due to a stronger negativity in `EMS-controlled' trials in AUGMENTED as compared to INVOLUNTARY. One possible explanation is that the trend is driven by false positive classifications, where a slight misalignment of the stimulation, i.e. it being too early, severely disrupts the user, eliciting a string prediction error signal. Taken together, we believe that contrasting and classifying single-trials could be a fruitful endeavor for approximating classification labels.

\subsection{Limitations \& Future Directions}
Two main procedural issues arose concerning the time estimation task: First, contrasting AUGMENTED with INTENTION, we noted that when EMS moved participants' fingers, they sometimes reported that their finger was pressing on the touchscreen for a longer duration as compared to their `normal' touchscreen tap. This may have introduced additional variance in the time estimation task since the exact moment of the tap that causes the tone is more obscure. Second, following pilot recordings, we chose to obtain participants' time estimates by asking them to type in their estimates on the keyboard. However, we observed that many participants, while perceiving a continuous distribution of the delays, did not answer at a continuous ms resolution but rather at steps of 50 or 100 ms, skewing the distribution of their answers. While many different versions of the intentional binding paradigm exist~\cite{Moore2012-dk}, we would choose a continuous slider for future experiments with different initial positions over trials to reduce the bias in participants' estimates.

\subsubsection{System Performance}
In the interviews, participants reported only a relatively low number of correctly detected intentions. This may have ultimately led to a misalignment between users' perceptions and the observed, measurable, system behavior. Furthermore, the necessity for a fixed block/condition order may have further contributed to this effect. However, keeping the order was necessary in order to first obtain training data for the BCI based on the unique EEG signals of each participant.

As a consequence, the insights gained into the system's potential to preserve a sense of agency are limited. The quantitative metrics employed in our study, specifically those measuring the sense of control (including temporal binding and item assessment), may not have captured the nuanced temporal alignment experiences associated with EMS augmentation and user movement intentions but rather a more holistic assessment of the entire experimental block, which included trials with missed stimulations, too early stimulations, and some trials where the stimulation was in alignment with the participants' intent.

Our system ran at a 10 Hz update rate. This was chosen in order to avoid buffer overflow due to the fact that the end-to-end computing latency from measuring the physiological signal to outputting the predicted class and probability took about 40 ms. The RP is reported to occur several hundred milliseconds prior to the physical movement~\cite{Schurger2021-vp}. Our augmentation then should have resulted in a motion pre-emption, considering a subtraction of the computing latency and a delay that the EMS stimulation takes to affect the muscle from the RP-based intention detection. However, it is difficult to assess whether participants' movements were in fact pre-empted as no readily available label exists for the `true' intention onset to compare the classifier to. That participants were able to act on their own volition, and hence no direct label of their movement intention could be derived, is the key distinction to related works using controlled stimulus-response paradigms.

While we believe that our prototype did not perform at a consistently high enough performance to allow for more fine-grained inferences about the the pre-emptive gain it achieved, we maintain that in some cases it elicited a pre-emption that maintained agency to a certain degree. We believe this to be a very promising finding because our prototype with low technological requirements still produces behavioral results indicating that intention can be detected and planned movements be augmented. However, from stimulus-response paradigms we recall that increasing participants' reaction times by about 80 ms is optimal with regards to maintaining SoA~\cite{Kasahara2019-sk}. On one hand, our system is capable of delivering pre-emption within this `SoA-optimal' time range. On the other hand, however, the sub-optimal performance of the classifier resulted in a lot of variation due to false positive detections. While in some cases, as we observed in the users' comments, the stimulation may have pre-empted a user's motion, false positive stimulation may have had a significant impact on the overall impression of the system.

Taken together, the \textit{key} objective remains in improving the overall performance of the classification system. This can be accomplished by fusing several complementary models. For virtual reality (VR) platforms,~\citet{David-John2021-vg} have recently shown that gaze dynamics, especially gaze velocity, carry information with respect to the intent to interact. Furthermore,~\citet{Nguyen2023-me} presented further evidence that physiological signals originating from muscle activity (EMG) provide a very reliable label of movement onset~\cite{Nguyen2023-me}. Arguably, an augmentation system based on several models tuned to specific physiological features will reach sufficient classification performance in the very near future.


\section{Conclusions}
In this paper, we designed and investigated a system to maintain users' SoA during augmented experiences using brain signals reflecting the intent to (inter-) act. In our user study, we found no convincing evidence that intentional binding effects are stronger when participants work with an augmentation system compared to being passively moved. However, participants rated their level of control working with the system higher than when being passively moved.

We believe this to be an important next step towards augmented users with full integration of the technology~\cite{Mueller2020-dl}. Ultimately, our closed-loop system is another step towards novel predictive interfaces that leverage the users' body directly yet still feel \textit{natural} as they align with the users' intention. For example, we envision that such closed-loop stimulation systems will prove useful in altering the affordance structure of an interaction in real-time~\cite{Gehrke2022-kz, Lopes2015-ze, Nataraj2020-wm}.

With AI technologies becoming more and more ubiquitous in our everyday lives, questions about how agency is shared between us humans and AI technologies arise regularly. With the likely future that these systems move directly onto our bodies, alignment with users' intentions will be the \textit{key} component driving their adoption.

%% file: input/9_acknowledgements.tex

ChatGPT (OpenAI, San Francisco, USA) was used to copy-edit author-generated content.

%% file: main.bbl

\begin{thebibliography}{68}


\ifx \showCODEN    \undefined \def \showCODEN     #1{\unskip}     \fi
\ifx \showDOI      \undefined \def \showDOI       #1{#1}\fi
\ifx \showISBNx    \undefined \def \showISBNx     #1{\unskip}     \fi
\ifx \showISBNxiii \undefined \def \showISBNxiii  #1{\unskip}     \fi
\ifx \showISSN     \undefined \def \showISSN      #1{\unskip}     \fi
\ifx \showLCCN     \undefined \def \showLCCN      #1{\unskip}     \fi
\ifx \shownote     \undefined \def \shownote      #1{#1}          \fi
\ifx \showarticletitle \undefined \def \showarticletitle #1{#1}   \fi
\ifx \showURL      \undefined \def \showURL       {\relax}        \fi
\providecommand\bibfield[2]{#2}
\providecommand\bibinfo[2]{#2}
\providecommand\natexlab[1]{#1}
\providecommand\showeprint[2][]{arXiv:#2}

\bibitem[Bates et~al\mbox{.}(2015)]%
        {Bates2015-bh}
\bibfield{author}{\bibinfo{person}{Douglas Bates}, \bibinfo{person}{Martin Mächler}, \bibinfo{person}{Ben Bolker}, {and} \bibinfo{person}{Steve Walker}.} \bibinfo{year}{2015}\natexlab{}.
\newblock \showarticletitle{Fitting Linear Mixed-Effects Models Using {lme4}}.
\newblock \bibinfo{journal}{\emph{J. Stat. Softw.}}  \bibinfo{volume}{67} (\bibinfo{date}{Oct.} \bibinfo{year}{2015}), \bibinfo{pages}{1--48}.
\newblock


\bibitem[Benjamini and Hochberg(1995)]%
        {Benjamini1995-cw}
\bibfield{author}{\bibinfo{person}{Yoav Benjamini} {and} \bibinfo{person}{Yosef Hochberg}.} \bibinfo{year}{1995}\natexlab{}.
\newblock \showarticletitle{Controlling the false discovery rate: A practical and powerful approach to multiple testing}.
\newblock \bibinfo{journal}{\emph{J. R. Stat. Soc.}} \bibinfo{volume}{57}, \bibinfo{number}{1} (\bibinfo{date}{Jan.} \bibinfo{year}{1995}), \bibinfo{pages}{289--300}.
\newblock


\bibitem[Berberian et~al\mbox{.}(2012)]%
        {Berberian2012-do}
\bibfield{author}{\bibinfo{person}{Bruno Berberian}, \bibinfo{person}{Jean-Christophe Sarrazin}, \bibinfo{person}{Patrick Le~Blaye}, {and} \bibinfo{person}{Patrick Haggard}.} \bibinfo{year}{2012}\natexlab{}.
\newblock \showarticletitle{Automation technology and sense of control: a window on human agency}.
\newblock \bibinfo{journal}{\emph{PLoS One}} \bibinfo{volume}{7}, \bibinfo{number}{3} (\bibinfo{date}{March} \bibinfo{year}{2012}), \bibinfo{pages}{e34075}.
\newblock


\bibitem[Bergström et~al\mbox{.}(2022)]%
        {Bergstrom2022-fb}
\bibfield{author}{\bibinfo{person}{Joanna Bergström}, \bibinfo{person}{Jarrod Knibbe}, \bibinfo{person}{Henning Pohl}, {and} \bibinfo{person}{Kasper Hornbæk}.} \bibinfo{year}{2022}\natexlab{}.
\newblock \showarticletitle{Sense of Agency and User Experience: Is There a Link?}
\newblock \bibinfo{journal}{\emph{ACM Trans. Comput.-Hum. Interact.}} \bibinfo{volume}{29}, \bibinfo{number}{4} (\bibinfo{date}{March} \bibinfo{year}{2022}), \bibinfo{pages}{1--22}.
\newblock


\bibitem[Bigdely-Shamlo et~al\mbox{.}(2015)]%
        {Bigdely-Shamlo2015-ds}
\bibfield{author}{\bibinfo{person}{Nima Bigdely-Shamlo}, \bibinfo{person}{Tim Mullen}, \bibinfo{person}{Christian Kothe}, \bibinfo{person}{Kyung-Min Su}, {and} \bibinfo{person}{Kay~A Robbins}.} \bibinfo{year}{2015}\natexlab{}.
\newblock \showarticletitle{The {PREP} pipeline: standardized preprocessing for large-scale {EEG} analysis}.
\newblock \bibinfo{journal}{\emph{Front. Neuroinform.}} \bibinfo{volume}{9}, \bibinfo{number}{June} (\bibinfo{date}{June} \bibinfo{year}{2015}), \bibinfo{pages}{16}.
\newblock


\bibitem[Blakemore et~al\mbox{.}(2002)]%
        {Blakemore2002-dj}
\bibfield{author}{\bibinfo{person}{Sarah~Jayne Blakemore}, \bibinfo{person}{Daniel~M Wolpert}, {and} \bibinfo{person}{Christopher~D Frith}.} \bibinfo{year}{2002}\natexlab{}.
\newblock \showarticletitle{Abnormalities in the awareness of action}.
\newblock \bibinfo{journal}{\emph{Trends Cogn. Sci.}} \bibinfo{volume}{6}, \bibinfo{number}{6} (\bibinfo{date}{June} \bibinfo{year}{2002}), \bibinfo{pages}{237--242}.
\newblock


\bibitem[Clark(2013)]%
        {Clark2013-ah}
\bibfield{author}{\bibinfo{person}{Andy Clark}.} \bibinfo{year}{2013}\natexlab{}.
\newblock \showarticletitle{Whatever next? Predictive brains, situated agents, and the future of cognitive science}.
\newblock \bibinfo{journal}{\emph{Behav. Brain Sci.}} \bibinfo{volume}{36}, \bibinfo{number}{3} (\bibinfo{date}{June} \bibinfo{year}{2013}), \bibinfo{pages}{181--204}.
\newblock


\bibitem[Cornelio et~al\mbox{.}(2022)]%
        {Cornelio2022-aq}
\bibfield{author}{\bibinfo{person}{Patricia Cornelio}, \bibinfo{person}{Patrick Haggard}, \bibinfo{person}{Kasper Hornbaek}, \bibinfo{person}{Orestis Georgiou}, \bibinfo{person}{Joanna Bergström}, \bibinfo{person}{Sriram Subramanian}, {and} \bibinfo{person}{Marianna Obrist}.} \bibinfo{year}{2022}\natexlab{}.
\newblock \showarticletitle{The sense of agency in emerging technologies for human-computer integration: A review}.
\newblock \bibinfo{journal}{\emph{Front. Neurosci.}}  \bibinfo{volume}{16} (\bibinfo{date}{Sept.} \bibinfo{year}{2022}), \bibinfo{pages}{949138}.
\newblock


\bibitem[Danry et~al\mbox{.}(2022)]%
        {Danry2022-xk}
\bibfield{author}{\bibinfo{person}{Valdemar Danry}, \bibinfo{person}{Pat Pataranutaporn}, \bibinfo{person}{Florian Mueller}, \bibinfo{person}{Pattie Maes}, {and} \bibinfo{person}{Sang-Won Leigh}.} \bibinfo{year}{2022}\natexlab{}.
\newblock \showarticletitle{On Eliciting a Sense of Self when Integrating with Computers}. In \bibinfo{booktitle}{\emph{Augmented Humans 2022}} \emph{(\bibinfo{series}{AHs 2022})}. \bibinfo{publisher}{Association for Computing Machinery}, \bibinfo{address}{New York, NY, USA}, \bibinfo{pages}{68--81}.
\newblock


\bibitem[David-John et~al\mbox{.}(2021)]%
        {David-John2021-vg}
\bibfield{author}{\bibinfo{person}{Brendan David-John}, \bibinfo{person}{Candace Peacock}, \bibinfo{person}{Ting Zhang}, \bibinfo{person}{T~Scott Murdison}, \bibinfo{person}{Hrvoje Benko}, {and} \bibinfo{person}{Tanya~R Jonker}.} \bibinfo{year}{2021}\natexlab{}.
\newblock \showarticletitle{Towards gaze-based prediction of the intent to interact in virtual reality}. In \bibinfo{booktitle}{\emph{ACM Symposium on Eye Tracking Research and Applications}} \emph{(\bibinfo{series}{ETRA '21 Short Papers}, \bibinfo{number}{Article 2})}. \bibinfo{publisher}{Association for Computing Machinery}, \bibinfo{address}{New York, NY, USA}, \bibinfo{pages}{1--7}.
\newblock


\bibitem[Deecke et~al\mbox{.}(1969)]%
        {Deecke1969-bl}
\bibfield{author}{\bibinfo{person}{L Deecke}, \bibinfo{person}{P Scheid}, {and} \bibinfo{person}{H~H Kornhuber}.} \bibinfo{year}{1969}\natexlab{}.
\newblock \showarticletitle{Distribution of readiness potential, pre-motion positivity, and motor potential of the human cerebral cortex preceding voluntary finger movements}.
\newblock \bibinfo{journal}{\emph{Exp. Brain Res.}} \bibinfo{volume}{7}, \bibinfo{number}{2} (\bibinfo{year}{1969}), \bibinfo{pages}{158--168}.
\newblock


\bibitem[Delorme and Makeig(2004)]%
        {Delorme2004-sn}
\bibfield{author}{\bibinfo{person}{Arnaud Delorme} {and} \bibinfo{person}{Scott Makeig}.} \bibinfo{year}{2004}\natexlab{}.
\newblock \showarticletitle{{EEGLAB}: an open source toolbox for analysis of single-trial {EEG} dynamics including independent component analysis}.
\newblock \bibinfo{journal}{\emph{J. Neurosci. Methods}} \bibinfo{volume}{134}, \bibinfo{number}{1} (\bibinfo{date}{March} \bibinfo{year}{2004}), \bibinfo{pages}{9--21}.
\newblock


\bibitem[Ebert and Wegner(2010)]%
        {Ebert2010-lu}
\bibfield{author}{\bibinfo{person}{Jeffrey~P Ebert} {and} \bibinfo{person}{Daniel~M Wegner}.} \bibinfo{year}{2010}\natexlab{}.
\newblock \showarticletitle{Time warp: authorship shapes the perceived timing of actions and events}.
\newblock \bibinfo{journal}{\emph{Conscious. Cogn.}} \bibinfo{volume}{19}, \bibinfo{number}{1} (\bibinfo{date}{March} \bibinfo{year}{2010}), \bibinfo{pages}{481--489}.
\newblock


\bibitem[Frith et~al\mbox{.}(2000)]%
        {Frith2000-ch}
\bibfield{author}{\bibinfo{person}{C~D Frith}, \bibinfo{person}{S~J Blakemore}, {and} \bibinfo{person}{D~M Wolpert}.} \bibinfo{year}{2000}\natexlab{}.
\newblock \showarticletitle{Abnormalities in the awareness and control of action}.
\newblock \bibinfo{journal}{\emph{Philos. Trans. R. Soc. Lond. B Biol. Sci.}} \bibinfo{volume}{355}, \bibinfo{number}{1404} (\bibinfo{date}{Dec.} \bibinfo{year}{2000}), \bibinfo{pages}{1771--1788}.
\newblock


\bibitem[Frith and Frith(2006)]%
        {Frith2006-sc}
\bibfield{author}{\bibinfo{person}{Chris~D Frith} {and} \bibinfo{person}{Uta Frith}.} \bibinfo{year}{2006}\natexlab{}.
\newblock \showarticletitle{How we predict what other people are going to do}.
\newblock \bibinfo{journal}{\emph{Brain Res.}} \bibinfo{volume}{1079}, \bibinfo{number}{1} (\bibinfo{date}{March} \bibinfo{year}{2006}), \bibinfo{pages}{36--46}.
\newblock


\bibitem[Gehrke et~al\mbox{.}(2019)]%
        {Gehrke2019-og}
\bibfield{author}{\bibinfo{person}{Lukas Gehrke}, \bibinfo{person}{Sezen Akman}, \bibinfo{person}{Pedro Lopes}, \bibinfo{person}{Albert Chen}, \bibinfo{person}{Avinash~Kumar Singh}, \bibinfo{person}{Hsiang-Ting Chen}, \bibinfo{person}{Chin-Teng Lin}, {and} \bibinfo{person}{Klaus Gramann}.} \bibinfo{year}{2019}\natexlab{}.
\newblock \showarticletitle{Detecting visuo-haptic mismatches in virtual reality using the prediction error negativity of event-related brain potentials}. In \bibinfo{booktitle}{\emph{Proceedings of the 2019 CHI Conference on Human Factors in Computing Systems - CHI '19}} \emph{(\bibinfo{series}{CHI '19})}. \bibinfo{publisher}{ACM Press}, \bibinfo{address}{New York, New York, USA}, \bibinfo{pages}{427:1--427:11}.
\newblock


\bibitem[Gehrke et~al\mbox{.}(2022)]%
        {Gehrke2022-kz}
\bibfield{author}{\bibinfo{person}{Lukas Gehrke}, \bibinfo{person}{Pedro Lopes}, {and} \bibinfo{person}{Klaus Gramann}.} \bibinfo{year}{2022}\natexlab{}.
\newblock \showarticletitle{Toward Human Augmentation Using Neural Fingerprints of Affordances}.
\newblock In \bibinfo{booktitle}{\emph{Affordances in Everyday Life: A Multidisciplinary Collection of Essays}}, \bibfield{editor}{\bibinfo{person}{Zakaria Djebbara}} (Ed.). \bibinfo{publisher}{Springer International Publishing}, \bibinfo{address}{Cham}, \bibinfo{pages}{173--180}.
\newblock


\bibitem[Gehrke et~al\mbox{.}(2024)]%
        {Gehrke2024-xq}
\bibfield{author}{\bibinfo{person}{Lukas Gehrke}, \bibinfo{person}{Leonie Terfurth}, \bibinfo{person}{Sezen Akman}, {and} \bibinfo{person}{Klaus Gramann}.} \bibinfo{year}{2024}\natexlab{}.
\newblock \showarticletitle{Visuo-haptic prediction errors: a multimodal dataset ({EEG}, motion) in {BIDS} format indexing mismatches in haptic interaction}.
\newblock \bibinfo{journal}{\emph{Front. Neuroergonomics}}  \bibinfo{volume}{5} (\bibinfo{date}{June} \bibinfo{year}{2024}), \bibinfo{pages}{1411305}.
\newblock


\bibitem[Gilbert et~al\mbox{.}(2019)]%
        {Gilbert2019-uc}
\bibfield{author}{\bibinfo{person}{F Gilbert}, \bibinfo{person}{M Cook}, \bibinfo{person}{T O'Brien}, {and} \bibinfo{person}{J Illes}.} \bibinfo{year}{2019}\natexlab{}.
\newblock \showarticletitle{Embodiment and Estrangement: Results from a First-in-Human “Intelligent {BCI”} Trial}.
\newblock \bibinfo{journal}{\emph{Sci. Eng. Ethics}} \bibinfo{volume}{25}, \bibinfo{number}{1} (\bibinfo{date}{Feb.} \bibinfo{year}{2019}), \bibinfo{pages}{83--96}.
\newblock


\bibitem[Gilbert et~al\mbox{.}(2017)]%
        {Gilbert2017-ze}
\bibfield{author}{\bibinfo{person}{Frederic Gilbert}, \bibinfo{person}{Eliza Goddard}, \bibinfo{person}{John Noel~M Viaña}, \bibinfo{person}{Adrian Carter}, {and} \bibinfo{person}{Malcolm Horne}.} \bibinfo{year}{2017}\natexlab{}.
\newblock \showarticletitle{{I} Miss Being Me: Phenomenological Effects of Deep Brain Stimulation}.
\newblock \bibinfo{journal}{\emph{AJOB Neurosci.}} \bibinfo{volume}{8}, \bibinfo{number}{2} (\bibinfo{date}{April} \bibinfo{year}{2017}), \bibinfo{pages}{96--109}.
\newblock


\bibitem[Goto et~al\mbox{.}(2020)]%
        {Goto2020-mw}
\bibfield{author}{\bibinfo{person}{Takashi Goto}, \bibinfo{person}{Swagata Das}, \bibinfo{person}{Katrin Wolf}, \bibinfo{person}{Pedro Lopes}, \bibinfo{person}{Yuichi Kurita}, {and} \bibinfo{person}{Kai Kunze}.} \bibinfo{year}{2020}\natexlab{}.
\newblock \showarticletitle{Accelerating Skill Acquisition of Two-Handed Drumming using Pneumatic Artificial Muscles}. In \bibinfo{booktitle}{\emph{Proceedings of the Augmented Humans International Conference}} \emph{(\bibinfo{series}{AHs '20}, \bibinfo{number}{Article 12})}. \bibinfo{publisher}{Association for Computing Machinery}, \bibinfo{address}{New York, NY, USA}, \bibinfo{pages}{1--9}.
\newblock


\bibitem[Gramfort et~al\mbox{.}(2013)]%
        {Gramfort2013-fa}
\bibfield{author}{\bibinfo{person}{Alexandre Gramfort}, \bibinfo{person}{Martin Luessi}, \bibinfo{person}{Eric Larson}, \bibinfo{person}{Denis~A Engemann}, \bibinfo{person}{Daniel Strohmeier}, \bibinfo{person}{Christian Brodbeck}, \bibinfo{person}{Roman Goj}, \bibinfo{person}{Mainak Jas}, \bibinfo{person}{Teon Brooks}, \bibinfo{person}{Lauri Parkkonen}, {and} \bibinfo{person}{Matti Hämäläinen}.} \bibinfo{year}{2013}\natexlab{}.
\newblock \showarticletitle{{MEG} and {EEG} data analysis with {MNE}-Python}.
\newblock \bibinfo{journal}{\emph{Front. Neurosci.}}  \bibinfo{volume}{7} (\bibinfo{date}{Dec.} \bibinfo{year}{2013}), \bibinfo{pages}{267}.
\newblock


\bibitem[Gutzeit et~al\mbox{.}(2023)]%
        {Gutzeit2023-ei}
\bibfield{author}{\bibinfo{person}{Julian Gutzeit}, \bibinfo{person}{Lisa Weller}, \bibinfo{person}{Jens Kürten}, {and} \bibinfo{person}{Lynn Huestegge}.} \bibinfo{year}{2023}\natexlab{}.
\newblock \showarticletitle{Intentional binding: Merely a procedural confound?}
\newblock \bibinfo{journal}{\emph{J. Exp. Psychol. Hum. Percept. Perform.}} \bibinfo{volume}{49}, \bibinfo{number}{6} (\bibinfo{date}{June} \bibinfo{year}{2023}), \bibinfo{pages}{759--773}.
\newblock


\bibitem[Haggard(2017)]%
        {Haggard2017-uv}
\bibfield{author}{\bibinfo{person}{Patrick Haggard}.} \bibinfo{year}{2017}\natexlab{}.
\newblock \showarticletitle{Sense of agency in the human brain}.
\newblock \bibinfo{journal}{\emph{Nat. Rev. Neurosci.}} \bibinfo{volume}{18}, \bibinfo{number}{4} (\bibinfo{date}{April} \bibinfo{year}{2017}), \bibinfo{pages}{196--207}.
\newblock


\bibitem[Haggard and Clark(2003)]%
        {Haggard2003-ff}
\bibfield{author}{\bibinfo{person}{Patrick Haggard} {and} \bibinfo{person}{Sam Clark}.} \bibinfo{year}{2003}\natexlab{}.
\newblock \showarticletitle{Intentional action: conscious experience and neural prediction}.
\newblock \bibinfo{journal}{\emph{Conscious. Cogn.}} \bibinfo{volume}{12}, \bibinfo{number}{4} (\bibinfo{date}{Dec.} \bibinfo{year}{2003}), \bibinfo{pages}{695--707}.
\newblock


\bibitem[Haggard et~al\mbox{.}(2002)]%
        {Haggard2002-sz}
\bibfield{author}{\bibinfo{person}{Patrick Haggard}, \bibinfo{person}{Sam Clark}, {and} \bibinfo{person}{Jeri Kalogeras}.} \bibinfo{year}{2002}\natexlab{}.
\newblock \showarticletitle{Voluntary action and conscious awareness}.
\newblock \bibinfo{journal}{\emph{Nat. Neurosci.}} \bibinfo{volume}{5}, \bibinfo{number}{4} (\bibinfo{date}{April} \bibinfo{year}{2002}), \bibinfo{pages}{382--385}.
\newblock


\bibitem[Hoerl et~al\mbox{.}(2020)]%
        {Hoerl2020-my}
\bibfield{author}{\bibinfo{person}{Christoph Hoerl}, \bibinfo{person}{Sara Lorimer}, \bibinfo{person}{Teresa McCormack}, \bibinfo{person}{David~A Lagnado}, \bibinfo{person}{Emma Blakey}, \bibinfo{person}{Emma~C Tecwyn}, {and} \bibinfo{person}{Marc~J Buehner}.} \bibinfo{year}{2020}\natexlab{}.
\newblock \showarticletitle{Temporal Binding, Causation, and Agency: Developing a New Theoretical Framework}.
\newblock \bibinfo{journal}{\emph{Cogn. Sci.}} \bibinfo{volume}{44}, \bibinfo{number}{5} (\bibinfo{date}{May} \bibinfo{year}{2020}), \bibinfo{pages}{e12843}.
\newblock


\bibitem[Jasper(1983)]%
        {Jasper1983-uw}
\bibfield{author}{\bibinfo{person}{H Jasper}.} \bibinfo{year}{1983}\natexlab{}.
\newblock \showarticletitle{The ten-twenty electrode system of the international federation, recommendations for the practice of clinical neurophysiology}.
\newblock \bibinfo{journal}{\emph{The International Federation Societies for Electroencephalography and Clinical Neurophysiology, Elsevier, Amsterdam}} (\bibinfo{year}{1983}), \bibinfo{pages}{3--10}.
\newblock


\bibitem[Kasahara et~al\mbox{.}(2019)]%
        {Kasahara2019-sk}
\bibfield{author}{\bibinfo{person}{Shunichi Kasahara}, \bibinfo{person}{Jun Nishida}, {and} \bibinfo{person}{Pedro Lopes}.} \bibinfo{year}{2019}\natexlab{}.
\newblock \showarticletitle{Preemptive Action: Accelerating Human Reaction using Electrical Muscle Stimulation Without Compromising Agency}. In \bibinfo{booktitle}{\emph{Proceedings of the 2019 CHI Conference on Human Factors in Computing Systems}} \emph{(\bibinfo{series}{CHI '19}, \bibinfo{number}{Paper 643})}. \bibinfo{publisher}{Association for Computing Machinery}, \bibinfo{address}{New York, NY, USA}, \bibinfo{pages}{1--15}.
\newblock


\bibitem[Kasahara et~al\mbox{.}(2021)]%
        {Kasahara2021-gy}
\bibfield{author}{\bibinfo{person}{Shunichi Kasahara}, \bibinfo{person}{Kazuma Takada}, \bibinfo{person}{Jun Nishida}, \bibinfo{person}{Kazuhisa Shibata}, \bibinfo{person}{Shinsuke Shimojo}, {and} \bibinfo{person}{Pedro Lopes}.} \bibinfo{year}{2021}\natexlab{}.
\newblock \showarticletitle{Preserving Agency During Electrical Muscle Stimulation Training Speeds up Reaction Time Directly After Removing {EMS}}. In \bibinfo{booktitle}{\emph{Proceedings of the 2021 CHI Conference on Human Factors in Computing Systems}} \emph{(\bibinfo{series}{CHI '21}, \bibinfo{number}{Article 194})}. \bibinfo{publisher}{Association for Computing Machinery}, \bibinfo{address}{New York, NY, USA}, \bibinfo{pages}{1--9}.
\newblock


\bibitem[Klug et~al\mbox{.}(2022)]%
        {Klug2022-lc}
\bibfield{author}{\bibinfo{person}{M Klug}, \bibinfo{person}{S Jeung}, \bibinfo{person}{A Wunderlich}, \bibinfo{person}{L Gehrke}, \bibinfo{person}{J Protzak}, \bibinfo{person}{Z Djebbara}, \bibinfo{person}{A Argubi-Wollesen}, \bibinfo{person}{B Wollesen}, {and} \bibinfo{person}{K Gramann}.} \bibinfo{year}{2022}\natexlab{}.
\newblock \showarticletitle{The {BeMoBIL} Pipeline for automated analyses of multimodal mobile brain and body imaging data}.
\newblock \bibinfo{journal}{\emph{bioRxiv}} (\bibinfo{date}{Oct.} \bibinfo{year}{2022}), \bibinfo{pages}{2022.09.29.510051}.
\newblock


\bibitem[Kunze et~al\mbox{.}(2017)]%
        {Kunze2017-co}
\bibfield{author}{\bibinfo{person}{Kai Kunze}, \bibinfo{person}{Kouta Minamizawa}, \bibinfo{person}{Stephan Lukosch}, \bibinfo{person}{Masahiko Inami}, {and} \bibinfo{person}{Jun Rekimoto}.} \bibinfo{year}{2017}\natexlab{}.
\newblock \showarticletitle{Superhuman sports: Applying human augmentation to physical exercise}.
\newblock \bibinfo{journal}{\emph{IEEE Pervasive Comput.}} \bibinfo{volume}{16}, \bibinfo{number}{2} (\bibinfo{date}{April} \bibinfo{year}{2017}), \bibinfo{pages}{14--17}.
\newblock


\bibitem[Kühn et~al\mbox{.}(2013)]%
        {Kuhn2013-ls}
\bibfield{author}{\bibinfo{person}{Simone Kühn}, \bibinfo{person}{Marcel Brass}, {and} \bibinfo{person}{Patrick Haggard}.} \bibinfo{year}{2013}\natexlab{}.
\newblock \showarticletitle{Feeling in control: Neural correlates of experience of agency}.
\newblock \bibinfo{journal}{\emph{Cortex}} \bibinfo{volume}{49}, \bibinfo{number}{7} (\bibinfo{year}{2013}), \bibinfo{pages}{1935--1942}.
\newblock


\bibitem[Ledoit and Wolf(2004)]%
        {Ledoit2004-bi}
\bibfield{author}{\bibinfo{person}{Olivier Ledoit} {and} \bibinfo{person}{Michael Wolf}.} \bibinfo{year}{2004}\natexlab{}.
\newblock \showarticletitle{A well-conditioned estimator for large-dimensional covariance matrices}.
\newblock \bibinfo{journal}{\emph{J. Multivar. Anal.}} \bibinfo{volume}{88}, \bibinfo{number}{2} (\bibinfo{date}{Feb.} \bibinfo{year}{2004}), \bibinfo{pages}{365--411}.
\newblock


\bibitem[Lenth et~al\mbox{.}(2020)]%
        {Lenth2020-xk}
\bibfield{author}{\bibinfo{person}{Russell Lenth}, \bibinfo{person}{Henrik Singmann}, \bibinfo{person}{Jonathon Love}, \bibinfo{person}{Paul Buerkner}, {and} \bibinfo{person}{Maxime Herve}.} \bibinfo{year}{2020}\natexlab{}.
\newblock \showarticletitle{Package ‘emmeans’}.
\newblock \bibinfo{journal}{\emph{R package version 1.4.6.}} \bibinfo{volume}{34}, \bibinfo{number}{1} (\bibinfo{year}{2020}), \bibinfo{pages}{216--221}.
\newblock


\bibitem[Libet et~al\mbox{.}(1983)]%
        {Libet1983-qu}
\bibfield{author}{\bibinfo{person}{B Libet}, \bibinfo{person}{E~W Wright, Jr}, {and} \bibinfo{person}{C~A Gleason}.} \bibinfo{year}{1983}\natexlab{}.
\newblock \showarticletitle{Preparation- or intention-to-act, in relation to pre-event potentials recorded at the vertex}.
\newblock \bibinfo{journal}{\emph{Electroencephalogr. Clin. Neurophysiol.}} \bibinfo{volume}{56}, \bibinfo{number}{4} (\bibinfo{date}{Oct.} \bibinfo{year}{1983}), \bibinfo{pages}{367--372}.
\newblock


\bibitem[Limerick et~al\mbox{.}(2014)]%
        {Limerick2014-un}
\bibfield{author}{\bibinfo{person}{Hannah Limerick}, \bibinfo{person}{David Coyle}, {and} \bibinfo{person}{James~W Moore}.} \bibinfo{year}{2014}\natexlab{}.
\newblock \showarticletitle{The experience of agency in human-computer interactions: a review}.
\newblock \bibinfo{journal}{\emph{Front. Hum. Neurosci.}} \bibinfo{volume}{8}, \bibinfo{number}{AUG} (\bibinfo{date}{Aug.} \bibinfo{year}{2014}), \bibinfo{pages}{643}.
\newblock


\bibitem[Lopes et~al\mbox{.}(2015a)]%
        {Lopes2015-dk}
\bibfield{author}{\bibinfo{person}{Pedro Lopes}, \bibinfo{person}{Alexandra Ion}, \bibinfo{person}{Willi Mueller}, \bibinfo{person}{Daniel Hoffmann}, \bibinfo{person}{Patrik Jonell}, {and} \bibinfo{person}{Patrick Baudisch}.} \bibinfo{year}{2015}\natexlab{a}.
\newblock \showarticletitle{Proprioceptive Interaction}.
\newblock In \bibinfo{booktitle}{\emph{Proceedings of the 33rd Annual ACM Conference on Human Factors in Computing Systems}}. Vol.~\bibinfo{volume}{2015-April}. \bibinfo{publisher}{Association for Computing Machinery}, \bibinfo{address}{New York, NY, USA}, \bibinfo{pages}{939--948}.
\newblock


\bibitem[Lopes et~al\mbox{.}(2015b)]%
        {Lopes2015-ze}
\bibfield{author}{\bibinfo{person}{Pedro Lopes}, \bibinfo{person}{Patrik Jonell}, {and} \bibinfo{person}{Patrick Baudisch}.} \bibinfo{year}{2015}\natexlab{b}.
\newblock \showarticletitle{Affordance++: Allowing Objects to Communicate Dynamic Use}.
\newblock In \bibinfo{booktitle}{\emph{Proceedings of the 33rd Annual ACM Conference on Human Factors in Computing Systems}}. \bibinfo{publisher}{Association for Computing Machinery}, \bibinfo{address}{New York, NY, USA}, \bibinfo{pages}{2515--2524}.
\newblock


\bibitem[Mayring(2015)]%
        {Mayring2015-pp}
\bibfield{author}{\bibinfo{person}{Philipp Mayring}.} \bibinfo{year}{2015}\natexlab{}.
\newblock \bibinfo{booktitle}{\emph{Qualitative Inhaltsanalyse: Grundlagen und Techniken}}.
\newblock \bibinfo{publisher}{Beltz}.
\newblock


\bibitem[Miller and Parasuraman(2007)]%
        {Miller2007-rb}
\bibfield{author}{\bibinfo{person}{Christopher~A Miller} {and} \bibinfo{person}{Raja Parasuraman}.} \bibinfo{year}{2007}\natexlab{}.
\newblock \showarticletitle{Designing for flexible interaction between humans and automation: delegation interfaces for supervisory control}.
\newblock \bibinfo{journal}{\emph{Hum. Factors}} \bibinfo{volume}{49}, \bibinfo{number}{1} (\bibinfo{date}{Feb.} \bibinfo{year}{2007}), \bibinfo{pages}{57--75}.
\newblock


\bibitem[Moore(2016)]%
        {Moore2016-ub}
\bibfield{author}{\bibinfo{person}{James~W Moore}.} \bibinfo{year}{2016}\natexlab{}.
\newblock \bibinfo{title}{What is the sense of agency and why does it matter?}
\newblock
\newblock


\bibitem[Moore et~al\mbox{.}(2012)]%
        {Moore2012-ic}
\bibfield{author}{\bibinfo{person}{J~W Moore}, \bibinfo{person}{D Middleton}, \bibinfo{person}{P Haggard}, {and} \bibinfo{person}{P~C Fletcher}.} \bibinfo{year}{2012}\natexlab{}.
\newblock \showarticletitle{Exploring implicit and explicit aspects of sense of agency}.
\newblock \bibinfo{journal}{\emph{Conscious. Cogn.}} \bibinfo{volume}{21}, \bibinfo{number}{4} (\bibinfo{date}{Dec.} \bibinfo{year}{2012}), \bibinfo{pages}{1748--1753}.
\newblock


\bibitem[Moore and Obhi(2012)]%
        {Moore2012-dk}
\bibfield{author}{\bibinfo{person}{James~W Moore} {and} \bibinfo{person}{Sukhvinder~S Obhi}.} \bibinfo{year}{2012}\natexlab{}.
\newblock \showarticletitle{Intentional binding and the sense of agency: a review}.
\newblock \bibinfo{journal}{\emph{Conscious. Cogn.}} \bibinfo{volume}{21}, \bibinfo{number}{1} (\bibinfo{date}{March} \bibinfo{year}{2012}), \bibinfo{pages}{546--561}.
\newblock


\bibitem[Mueller et~al\mbox{.}(2020)]%
        {Mueller2020-dl}
\bibfield{author}{\bibinfo{person}{Florian~Floyd Mueller}, \bibinfo{person}{Pedro Lopes}, \bibinfo{person}{Paul Strohmeier}, \bibinfo{person}{Wendy Ju}, \bibinfo{person}{Caitlyn Seim}, \bibinfo{person}{Martin Weigel}, \bibinfo{person}{Suranga Nanayakkara}, \bibinfo{person}{Marianna Obrist}, \bibinfo{person}{Zhuying Li}, \bibinfo{person}{Joseph Delfa}, \bibinfo{person}{Jun Nishida}, \bibinfo{person}{Elizabeth~M Gerber}, \bibinfo{person}{Dag Svanaes}, \bibinfo{person}{Jonathan Grudin}, \bibinfo{person}{Stefan Greuter}, \bibinfo{person}{Kai Kunze}, \bibinfo{person}{Thomas Erickson}, \bibinfo{person}{Steven Greenspan}, \bibinfo{person}{Masahiko Inami}, \bibinfo{person}{Joe Marshall}, \bibinfo{person}{Harald Reiterer}, \bibinfo{person}{Katrin Wolf}, \bibinfo{person}{Jochen Meyer}, \bibinfo{person}{Thecla Schiphorst}, \bibinfo{person}{Dakuo Wang}, {and} \bibinfo{person}{Pattie Maes}.} \bibinfo{year}{2020}\natexlab{}.
\newblock \showarticletitle{Next steps for human-computer integration}. In \bibinfo{booktitle}{\emph{Proceedings of the 2020 CHI Conference on Human Factors in Computing Systems}}. \bibinfo{publisher}{ACM}, \bibinfo{address}{New York, NY, USA}, \bibinfo{pages}{1--15}.
\newblock


\bibitem[Nataraj et~al\mbox{.}(2020)]%
        {Nataraj2020-wm}
\bibfield{author}{\bibinfo{person}{Raviraj Nataraj}, \bibinfo{person}{Sean Sanford}, \bibinfo{person}{Mingxiao Liu}, \bibinfo{person}{Kevin Walsh}, \bibinfo{person}{Samuel Wilder}, \bibinfo{person}{Anthony Santo}, {and} \bibinfo{person}{David Hollinger}.} \bibinfo{year}{2020}\natexlab{}.
\newblock \showarticletitle{Cognitive and Physiological Intent for the Adaptation of Motor Prostheses}.
\newblock In \bibinfo{booktitle}{\emph{Advances in Motor Neuroprostheses}}, \bibfield{editor}{\bibinfo{person}{Ramana Vinjamuri}} (Ed.). \bibinfo{publisher}{Springer International Publishing}, \bibinfo{address}{Cham}, \bibinfo{pages}{123--153}.
\newblock


\bibitem[Nguyen et~al\mbox{.}(2023)]%
        {Nguyen2023-me}
\bibfield{author}{\bibinfo{person}{Willy Nguyen}, \bibinfo{person}{Klaus Gramann}, {and} \bibinfo{person}{Lukas Gehrke}.} \bibinfo{year}{2023}\natexlab{}.
\newblock \showarticletitle{Modeling the Intent to Interact with {VR} using Physiological Features}.
\newblock \bibinfo{journal}{\emph{IEEE Trans. Vis. Comput. Graph.}}  \bibinfo{volume}{PP} (\bibinfo{date}{Aug.} \bibinfo{year}{2023}).
\newblock


\bibitem[Niso et~al\mbox{.}(2023)]%
        {Niso2023-ce}
\bibfield{author}{\bibinfo{person}{Guiomar Niso}, \bibinfo{person}{Elena Romero}, \bibinfo{person}{Jeremy~T Moreau}, \bibinfo{person}{Alvaro Araujo}, {and} \bibinfo{person}{Laurens~R Krol}.} \bibinfo{year}{2023}\natexlab{}.
\newblock \showarticletitle{Wireless {EEG}: A survey of systems and studies}.
\newblock \bibinfo{journal}{\emph{Neuroimage}}  \bibinfo{volume}{269} (\bibinfo{date}{April} \bibinfo{year}{2023}), \bibinfo{pages}{119774}.
\newblock


\bibitem[Palmer et~al\mbox{.}(2011)]%
        {Palmer2011-zs}
\bibfield{author}{\bibinfo{person}{Jason Palmer}, \bibinfo{person}{Ken Kreutz-Delgado}, {and} \bibinfo{person}{Scott Makeig}.} \bibinfo{year}{2011}\natexlab{}.
\newblock \showarticletitle{{AMICA}: An Adaptive Mixture of Independent Component Analyzers with Shared Components}.
\newblock \bibinfo{journal}{\emph{San Diego, CA: Technical report, Swartz Center for Computational Neuroscience}} (\bibinfo{year}{2011}), \bibinfo{pages}{1--15}.
\newblock


\bibitem[Pangratz et~al\mbox{.}(2023)]%
        {Pangratz2023-ew}
\bibfield{author}{\bibinfo{person}{Elisabeth Pangratz}, \bibinfo{person}{Francesco Chiossi}, \bibinfo{person}{Steeven Villa}, \bibinfo{person}{Klaus Gramann}, {and} \bibinfo{person}{Lukas Gehrke}.} \bibinfo{year}{2023}\natexlab{}.
\newblock \showarticletitle{Towards an Implicit Metric of Sensory-Motor Accuracy: Brain Responses to Auditory Prediction Errors in Pianists}. In \bibinfo{booktitle}{\emph{Proceedings of the 15th Conference on Creativity and Cognition}} \emph{(\bibinfo{series}{C\&C '23})}. \bibinfo{publisher}{Association for Computing Machinery}, \bibinfo{address}{New York, NY, USA}, \bibinfo{pages}{129--138}.
\newblock


\bibitem[Papenmeier et~al\mbox{.}(2022)]%
        {Papenmeier2022-oi}
\bibfield{author}{\bibinfo{person}{Andrea Papenmeier}, \bibinfo{person}{Dagmar Kern}, \bibinfo{person}{Daniel Hienert}, \bibinfo{person}{Yvonne Kammerer}, {and} \bibinfo{person}{Christin Seifert}.} \bibinfo{year}{2022}\natexlab{}.
\newblock \showarticletitle{How Accurate Does It Feel? – Human Perception of Different Types of Classification Mistakes}. In \bibinfo{booktitle}{\emph{Proceedings of the 2022 CHI Conference on Human Factors in Computing Systems}} \emph{(\bibinfo{series}{CHI '22}, \bibinfo{number}{Article 180})}. \bibinfo{publisher}{Association for Computing Machinery}, \bibinfo{address}{New York, NY, USA}, \bibinfo{pages}{1--13}.
\newblock


\bibitem[Parés-Pujolràs et~al\mbox{.}(2019)]%
        {Pares-Pujolras2019-ll}
\bibfield{author}{\bibinfo{person}{Elisabeth Parés-Pujolràs}, \bibinfo{person}{Yong-Wook Kim}, \bibinfo{person}{Chang-Hwan Im}, {and} \bibinfo{person}{Patrick Haggard}.} \bibinfo{year}{2019}\natexlab{}.
\newblock \showarticletitle{Latent awareness: Early conscious access to motor preparation processes is linked to the readiness potential}.
\newblock \bibinfo{journal}{\emph{Neuroimage}}  \bibinfo{volume}{202} (\bibinfo{date}{Nov.} \bibinfo{year}{2019}), \bibinfo{pages}{116140}.
\newblock


\bibitem[Pedregosa et~al\mbox{.}(2012)]%
        {Pedregosa2012-sj}
\bibfield{author}{\bibinfo{person}{Fabian Pedregosa}, \bibinfo{person}{Gaël Varoquaux}, \bibinfo{person}{Alexandre Gramfort}, \bibinfo{person}{Vincent Michel}, \bibinfo{person}{Bertrand Thirion}, \bibinfo{person}{Olivier Grisel}, \bibinfo{person}{Mathieu Blondel}, \bibinfo{person}{Andreas Müller}, \bibinfo{person}{Joel Nothman}, \bibinfo{person}{Gilles Louppe}, \bibinfo{person}{Peter Prettenhofer}, \bibinfo{person}{Ron Weiss}, \bibinfo{person}{Vincent Dubourg}, \bibinfo{person}{Jake Vanderplas}, \bibinfo{person}{Alexandre Passos}, \bibinfo{person}{David Cournapeau}, \bibinfo{person}{Matthieu Brucher}, \bibinfo{person}{Matthieu Perrot}, {and} \bibinfo{person}{Édouard Duchesnay}.} \bibinfo{year}{2012}\natexlab{}.
\newblock \showarticletitle{Scikit-learn: Machine Learning in Python}.
\newblock \bibinfo{journal}{\emph{arXiv [cs.LG]}} (\bibinfo{date}{Jan.} \bibinfo{year}{2012}).
\newblock


\bibitem[Pinheiro and Bates(2006)]%
        {Pinheiro2006-bk}
\bibfield{author}{\bibinfo{person}{José Pinheiro} {and} \bibinfo{person}{Douglas Bates}.} \bibinfo{year}{2006}\natexlab{}.
\newblock \bibinfo{booktitle}{\emph{Mixed-Effects Models in {S} and {S}-{PLUS}}}.
\newblock \bibinfo{publisher}{Springer Science \& Business Media}.
\newblock


\bibitem[Pion-Tonachini et~al\mbox{.}(2019)]%
        {Pion-Tonachini2019-fy}
\bibfield{author}{\bibinfo{person}{Luca Pion-Tonachini}, \bibinfo{person}{Ken Kreutz-Delgado}, {and} \bibinfo{person}{Scott Makeig}.} \bibinfo{year}{2019}\natexlab{}.
\newblock \showarticletitle{{ICLabel}: An automated electroencephalographic independent component classifier, dataset, and website}.
\newblock \bibinfo{journal}{\emph{Neuroimage}}  \bibinfo{volume}{198} (\bibinfo{date}{Sept.} \bibinfo{year}{2019}), \bibinfo{pages}{181--197}.
\newblock


\bibitem[Schultze-Kraft et~al\mbox{.}(2016)]%
        {Schultze-Kraft2016-bx}
\bibfield{author}{\bibinfo{person}{Matthias Schultze-Kraft}, \bibinfo{person}{Daniel Birman}, \bibinfo{person}{Marco Rusconi}, \bibinfo{person}{Carsten Allefeld}, \bibinfo{person}{Kai Görgen}, \bibinfo{person}{Sven Dähne}, \bibinfo{person}{Benjamin Blankertz}, {and} \bibinfo{person}{John-Dylan Haynes}.} \bibinfo{year}{2016}\natexlab{}.
\newblock \showarticletitle{The point of no return in vetoing self-initiated movements}.
\newblock \bibinfo{journal}{\emph{Proc. Natl. Acad. Sci. U. S. A.}} \bibinfo{volume}{113}, \bibinfo{number}{4} (\bibinfo{date}{Jan.} \bibinfo{year}{2016}), \bibinfo{pages}{1080--1085}.
\newblock


\bibitem[Schultze-Kraft et~al\mbox{.}(2021)]%
        {Schultze-Kraft2021-cu}
\bibfield{author}{\bibinfo{person}{Matthias Schultze-Kraft}, \bibinfo{person}{Vincent Jonany}, \bibinfo{person}{Thomas~Samuel Binns}, \bibinfo{person}{Joram Soch}, \bibinfo{person}{Benjamin Blankertz}, {and} \bibinfo{person}{John-Dylan Haynes}.} \bibinfo{year}{2021}\natexlab{}.
\newblock \showarticletitle{Suppress Me if You Can: Neurofeedback of the Readiness Potential}.
\newblock \bibinfo{journal}{\emph{eNeuro}} \bibinfo{volume}{8}, \bibinfo{number}{2} (\bibinfo{date}{March} \bibinfo{year}{2021}), \bibinfo{pages}{1--11}.
\newblock


\bibitem[Schultze-Kraft et~al\mbox{.}(2020)]%
        {Schultze-Kraft2020-rm}
\bibfield{author}{\bibinfo{person}{Matthias Schultze-Kraft}, \bibinfo{person}{Elisabeth Parés-Pujolràs}, \bibinfo{person}{Karla Matić}, \bibinfo{person}{Patrick Haggard}, {and} \bibinfo{person}{John-Dylan Haynes}.} \bibinfo{year}{2020}\natexlab{}.
\newblock \showarticletitle{Preparation and execution of voluntary action both contribute to awareness of intention}.
\newblock \bibinfo{journal}{\emph{Proc. Biol. Sci.}} \bibinfo{volume}{287}, \bibinfo{number}{1923} (\bibinfo{date}{March} \bibinfo{year}{2020}), \bibinfo{pages}{20192928}.
\newblock


\bibitem[Schurger et~al\mbox{.}(2021)]%
        {Schurger2021-vp}
\bibfield{author}{\bibinfo{person}{Aaron Schurger}, \bibinfo{person}{Pengbo~'ben' Hu}, \bibinfo{person}{Joanna Pak}, {and} \bibinfo{person}{Adina~L Roskies}.} \bibinfo{year}{2021}\natexlab{}.
\newblock \showarticletitle{What Is the Readiness Potential?}
\newblock \bibinfo{journal}{\emph{Trends Cogn. Sci.}} \bibinfo{volume}{25}, \bibinfo{number}{7} (\bibinfo{date}{July} \bibinfo{year}{2021}), \bibinfo{pages}{558--570}.
\newblock


\bibitem[Shibasaki and Hallett(2006)]%
        {Shibasaki2006-mt}
\bibfield{author}{\bibinfo{person}{Hiroshi Shibasaki} {and} \bibinfo{person}{Mark Hallett}.} \bibinfo{year}{2006}\natexlab{}.
\newblock \showarticletitle{What is the bereitschaftspotential?}
\newblock \bibinfo{journal}{\emph{Clin. Neurophysiol.}} \bibinfo{volume}{117}, \bibinfo{number}{11} (\bibinfo{date}{Nov.} \bibinfo{year}{2006}), \bibinfo{pages}{2341--2356}.
\newblock


\bibitem[Si-mohammed et~al\mbox{.}(2020)]%
        {Si-mohammed2020-ru}
\bibfield{author}{\bibinfo{person}{Hakim Si-mohammed}, \bibinfo{person}{Catarina Lopes-dias}, \bibinfo{person}{Maria Duarte}, \bibinfo{person}{Camille Jeunet}, {and} \bibinfo{person}{Reinhold Scherer}.} \bibinfo{year}{2020}\natexlab{}.
\newblock \showarticletitle{Detecting System Errors in Virtual Reality Using {EEG} Through Error-Related Potentials}.
\newblock  \bibinfo{number}{1} (\bibinfo{year}{2020}), \bibinfo{pages}{653--661}.
\newblock


\bibitem[Suzuki et~al\mbox{.}(2019)]%
        {Suzuki2019-pi}
\bibfield{author}{\bibinfo{person}{Keisuke Suzuki}, \bibinfo{person}{Peter Lush}, \bibinfo{person}{Anil~K Seth}, {and} \bibinfo{person}{Warrick Roseboom}.} \bibinfo{year}{2019}\natexlab{}.
\newblock \showarticletitle{Intentional Binding Without Intentional Action}.
\newblock \bibinfo{journal}{\emph{Psychol. Sci.}} \bibinfo{volume}{30}, \bibinfo{number}{6} (\bibinfo{date}{June} \bibinfo{year}{2019}), \bibinfo{pages}{842--853}.
\newblock


\bibitem[Terfurth et~al\mbox{.}(2024)]%
        {Terfurth2024-kh}
\bibfield{author}{\bibinfo{person}{Leonie Terfurth}, \bibinfo{person}{Klaus Gramann}, {and} \bibinfo{person}{Lukas Gehrke}.} \bibinfo{year}{2024}\natexlab{}.
\newblock \showarticletitle{Decoding Realism of Virtual Objects: Exploring Behavioral and Ocular Reactions to Inaccurate Interaction Feedback}.
\newblock \bibinfo{journal}{\emph{ACM Trans. Comput.-Hum. Interact.}} (\bibinfo{date}{April} \bibinfo{year}{2024}).
\newblock


\bibitem[Travers et~al\mbox{.}(2020)]%
        {Travers2020-hf}
\bibfield{author}{\bibinfo{person}{Eoin Travers}, \bibinfo{person}{Nima Khalighinejad}, \bibinfo{person}{Aaron Schurger}, {and} \bibinfo{person}{Patrick Haggard}.} \bibinfo{year}{2020}\natexlab{}.
\newblock \showarticletitle{Do readiness potentials happen all the time?}
\newblock \bibinfo{journal}{\emph{Neuroimage}}  \bibinfo{volume}{206} (\bibinfo{date}{Feb.} \bibinfo{year}{2020}), \bibinfo{pages}{116286}.
\newblock


\bibitem[Tukey(1949)]%
        {Tukey1949-sl}
\bibfield{author}{\bibinfo{person}{J~W Tukey}.} \bibinfo{year}{1949}\natexlab{}.
\newblock \showarticletitle{Comparing individual means in the analysis of variance}.
\newblock \bibinfo{journal}{\emph{Biometrics}} \bibinfo{volume}{5}, \bibinfo{number}{2} (\bibinfo{date}{June} \bibinfo{year}{1949}), \bibinfo{pages}{99--114}.
\newblock


\bibitem[Wen and Haggard(2020)]%
        {Wen2020-dk}
\bibfield{author}{\bibinfo{person}{Wen Wen} {and} \bibinfo{person}{Patrick Haggard}.} \bibinfo{year}{2020}\natexlab{}.
\newblock \showarticletitle{Prediction error and regularity detection underlie two dissociable mechanisms for computing the sense of agency}.
\newblock \bibinfo{journal}{\emph{Cognition}}  \bibinfo{volume}{195} (\bibinfo{date}{Feb.} \bibinfo{year}{2020}), \bibinfo{pages}{104074}.
\newblock


\bibitem[Wen and Imamizu(2022)]%
        {Wen2022-bm}
\bibfield{author}{\bibinfo{person}{Wen Wen} {and} \bibinfo{person}{Hiroshi Imamizu}.} \bibinfo{year}{2022}\natexlab{}.
\newblock \showarticletitle{The sense of agency in perception, behaviour and human–machine interactions}.
\newblock \bibinfo{journal}{\emph{Nature Reviews Psychology}} \bibinfo{volume}{1}, \bibinfo{number}{4} (\bibinfo{date}{Feb.} \bibinfo{year}{2022}), \bibinfo{pages}{211--222}.
\newblock


\bibitem[Zander et~al\mbox{.}(2016)]%
        {Zander2016-ed}
\bibfield{author}{\bibinfo{person}{Thorsten~O Zander}, \bibinfo{person}{Laurens~R Krol}, \bibinfo{person}{Niels~P Birbaumer}, {and} \bibinfo{person}{Klaus Gramann}.} \bibinfo{year}{2016}\natexlab{}.
\newblock \showarticletitle{Neuroadaptive technology enables implicit cursor control based on medial prefrontal cortex activity}.
\newblock \bibinfo{journal}{\emph{Proc. Natl. Acad. Sci. U. S. A.}} \bibinfo{volume}{113}, \bibinfo{number}{52} (\bibinfo{date}{Dec.} \bibinfo{year}{2016}), \bibinfo{pages}{14898--14903}.
\newblock


\end{thebibliography}
